\documentclass[pra,twocolumn]{revtex4-1}
\usepackage{hyperref} 

\usepackage{graphicx} 
\usepackage{subfig}
\usepackage{color}
\usepackage{filecontents}
\usepackage{soul}
\usepackage{mathrsfs}

\newcommand{\ket}[1]{\left| #1 \right\rangle}
\newcommand{\bra}[1]{\left\langle #1 \right|}

\newcommand{\be}{\begin{equation}}
\newcommand{\ee}{\end{equation}}
\newcommand{\bea}{\begin{eqnarray}}
\newcommand{\eea}{\end{eqnarray}}
\newcommand*{\myeqref}[2][Eq.~]{%
\hyperref[{#2}]{#1(\ref*{#2})}%
}
\def\equationautorefname#1#2\null{%
Eq.#1(#2\null)%
}

\usepackage{dcolumn}
\usepackage{bm}
\usepackage{amssymb,amsmath}
\usepackage{color}
\usepackage{float}
\usepackage{tikz}
\usetikzlibrary{arrows}

\definecolor{DarkGreen}{rgb}{0,0.6,0.2}

\begin{document}
\title{Collective photon routing improvement in a dissipative quantum emitter chain strongly coupled to a chiral waveguide QED ladder}

\author{Bibandhan Poudyal and Imran M. Mirza}
\affiliation{Macklin Quantum Information Sciences, Department of Physics, Miami University, Oxford Ohio 45056, USA}
\email{mirzaim@miamioh.edu}

\begin{abstract}
We examine the routing scheme of single photons in a one-dimensional periodic chain of two-level quantum emitters (QEs) strongly coupled to two waveguides in a ladder configuration. It is known that for a single-emitter chiral waveguide ladder setting, photons can be redirected from one waveguide to another with a 100\% probability (deterministically) provided the resonance condition is met and spontaneous emission is completely ignored. However, when the spontaneous emission is included the routing scheme becomes considerably imperfect. In this paper, we present a solution to this issue by considering a chain of QEs where in addition to the waveguide mediated interaction among emitters, a direct and infinitely long-ranged dipole-dipole interaction (DDI) is taken into account. We show that the collective effects arising from the strong DDI protect the routing scheme from spontaneous emission loss. In particular, we demonstrate that the router operation can be improved from 58\% to $\sim$95\% in a typical dissipative chiral light-matter interface consisting of nanowires modes strongly interacting with a linear chain of 30 quantum dots. With the recent experimental progress in chiral quantum optics, trapped QEs evanescently coupled to tapered nanofibers can serve as a platform for the experimental realization of this work.
\end{abstract}

\maketitle

\section{Introduction}
Several quantum networking protocols rely on using single photons as the carrier of quantum information as they propagate fast and can cover long distances without considerably influenced by the decoherence effects \cite{kimble2008quantum}. Typically stationary qubits (quantum emitters) are placed at the nodes in these networks to store, manipulate, and retrieve the quantum information while the photons, guided by optical fibers, transfer information from one node to another. Switching or routing of single photons turns out to be a key requirement in such networks to interlink different nodes \cite{wilk2007single, ritter2012elementary}. For a reliable interlinking protocol, the routing scheme is required to be not only efficient but also able to preserve the quantum state of the photons initially launched into the network. 

Historically Harris and Yamamoto were the first to propose in 1998 that quantum interference can be utilized for photon switching in a four-level atomic system \cite{harris1998photon}. Afterward, in the last decade or so, several proposals have been put forward and experiments have been performed to accomplish efficient photon routing. For instance, it has been theoretically proposed that electromagnetically induced transparency (EIT) in cavity optomechanics can be exploited for weak probe field photon switching \cite{agarwal2012optomechanical}. There has also been a series of routing proposals reported which are based on QE embedded in coupled resonators waveguides (CRWs) (\cite{zhou2013quantum, lu2014single, lu2015t, qin2016controllable, ahumada2019tunable} to name a few). In some other proposals circuit QED systems have been used to attain routing through EIT (see for example \cite{xia2018quantum}). On the experimental side, photon switching or routing has also been demonstrated in cavity quantum electrodynamics (QED) setups \cite{aoki2009efficient, shomroni2014all} and in the microwave domain using artificial atoms (such as transmon qubit) interacting with transmission lines \cite{hoi2011demonstration}. 

Waveguide QED architectures are particularly appealing in this regard as these structures can be used to perform several quantum information tasks in a single setup \cite{roy2017colloquium} including the photon routing \cite{li2015designable, zhu2019single, yang2018phase}. Closely akin to waveguide QED is the burgeoning development of chiral quantum optical systems \cite{lodahl2017chiral}. Chiral atom-field interaction here refers to propagation-direction-dependent absorption/emission of light which is achieved by the transverse confinement of electromagnetic radiation in sub-wavelength diameter optical fibers. It is known that such confinement leads to the spin-momentum locking of light \cite{petersen2014chiral}. One interesting consequence of this effect is if we bring a QE close to such a fiber then if the polarization of light is matched with the polarization-dependent absorption of the QE the light propagation is locked with the absorption properties of the emitter. As a result, light propagating in a certain direction in the fiber can be absorbed by such an emitter but the light propagating in the opposite direction fails to be absorbed. Since 2012, several experiments have been performed on the subject of chiral quantum optics with many exciting applications \cite{mirza2016multiqubit, gonzalez2015chiral, mirza2016two, mahmoodian2017engineering, mirza2017chirality, mahmoodian2019dynamics, mirza2018influence, mirza2020dimer}. Today more than 90\% directionality and 98\% atom-waveguide coupling strength have been reported in photonic crystal waveguides \cite{sollner2015deterministic}.

\begin{figure*}
\includegraphics[width=6.5in,height=2.8in]{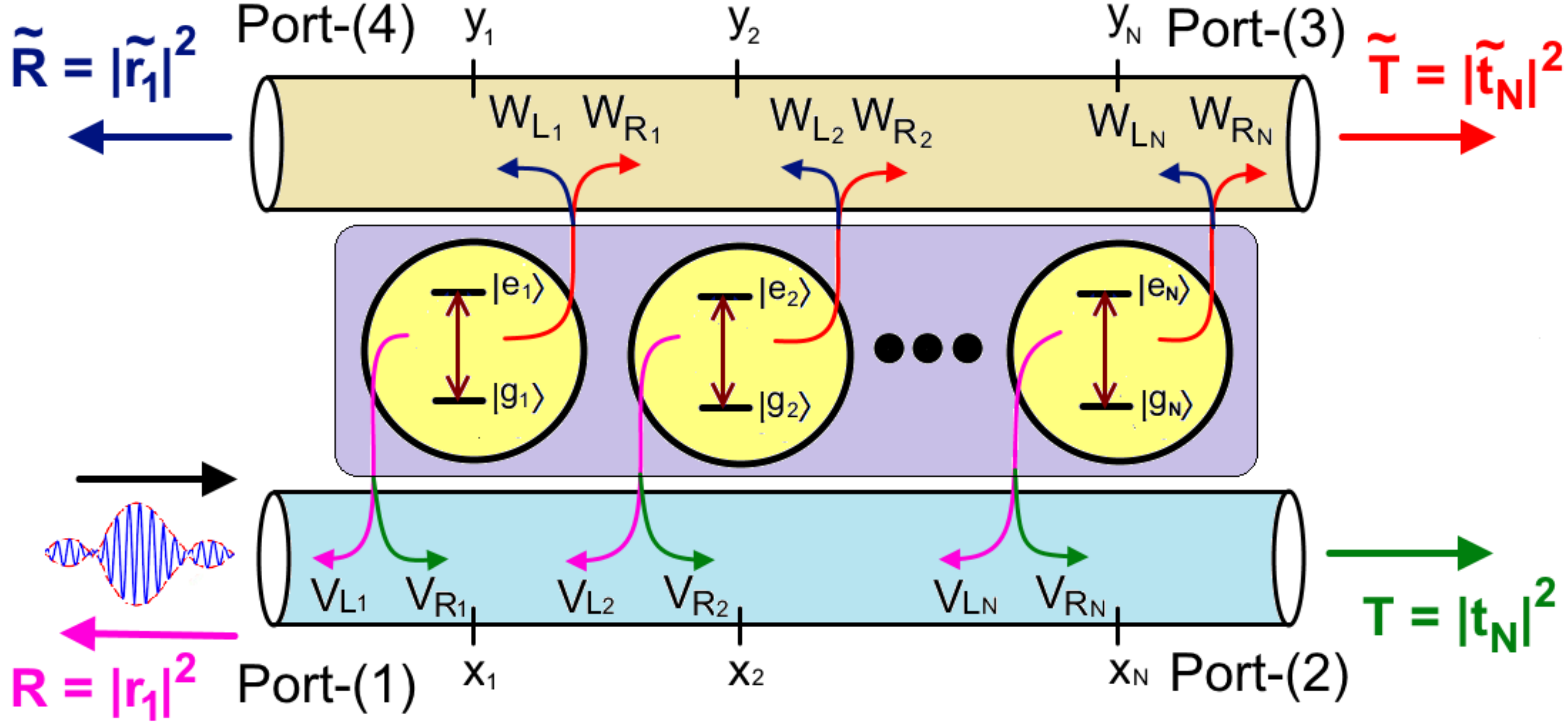}
\captionsetup{
 format=plain,
 margin=1em,
 justification=raggedright,
 singlelinecheck=false
}
\caption{(Color online) Illustration of a periodic multi-emitter chain coupled to a double waveguide QED system. A single-photon wavepacket is launched from the port-(1) and we are interested in port-(1) to port-(3) routing scheme. The purple region shows the presence of infinitely long-range dipole-dipole interaction among all QEs. $R$, $T$, $\widetilde{R}$, and $\widetilde{T}$ represent intensities of reflection from the bottom waveguide, transmission from the bottom waveguide, leftward rectification from the upper waveguide, and rightward rectification from the upper waveguide, respectively.}
\label{Fig1}
\end{figure*}

In this work, we analyze the nonreciprocal routing of single photons in many-emitter chiral waveguide QED in a ladder geometry. In such a four-port device, it is known that for a single-emitter case, perfect/deterministic routing can be achieved if the resonant interaction between the emitter and chiral waveguides is ensured in the absence of spontaneous emission (see for example works of Cheng et al. \cite{cheng2016single}, Gonzalez et al. \cite{gonzalez2016nonreciprocal}, and Yan et al. \cite{yan2018targeted}). However, in a more realistic scenario in which a finite spontaneous emission is included, routing efficiency is considerably impacted and shows a drastic reduction. Herein, we focus on this issue and discuss how the routing scheme can be notably improved in the presence of spontaneous emission loss. To this end, we consider a 1D chain of two-level QEs strongly coupled to the waveguide ladder setting. In our model, we further assume that in addition to waveguide mediated indirect interactions among QEs there also exists an infinitely long-ranged direct dipole-dipole interaction (DDI). To meet the DDI condition we suppose the inter-emitter separation is less than a single resonant wavelength \cite{milonni1974retardation}.

For the case of two dissipative QEs strongly coupled to both chiral waveguide, we analytically find that the output spectra split into a frequency doublet due to waveguide mediated interaction without DDI. The inclusion of DDI introduces an asymmetry in the peak heights which can be exploited to improve the routing efficiency in the presence of dissipation (spontaneous emission). Furthermore, we numerically extend our results to the many-emitter regime. Therein, we demonstrate that for a chain of 30 identical QEs strongly coupled through waveguide modes as well as through DDI, with an interaction rate $\Gamma=11.03\Gamma_0$ ($\Gamma_0$ being the free space decay rate) the routing efficiency can be improved from 58\% to $\sim$95\% with a DDI of strength $\sim 2.1\Gamma$ in the presence of spontaneous emission rate of $\sim 0.62\Gamma$.

The paper is organized as follows. In Sec. II we describe the system model. In Sec. III we discuss the single-photon transport theory in a many-emitter waveguide QED ladder. Sec. IV presents the results mainly focusing on the issue of how collective effects can promote routing efficiency. Finally, in Sec. V we summarize the main conclusions of this work.


\section{Model and Hamiltonian}
As shown in Fig.~\ref{Fig1}, we consider a periodic array of two-level QEs (herein also referred to qubits/atoms) simultaneously side coupled to two one-dimensional waveguides forming a double-waveguide multi-emitter QED ladder. The total Hamiltonian of the system can be decomposed into four pieces
\begin{equation}
\label{Hamil}
{\bf \hat{\mathscr{H}}}=\hat{\mathcal{H}}_{\rm 2LS}+\hat{\mathcal{H}}_{\rm wav}+\hat{\mathcal{H}}_{\rm int}+\hat{\mathcal{H}}_{\rm DDI}.
\end{equation}
$\hat{\mathcal{H}}_{\rm 2LS}$ describes the free-Hamiltonian of $N$ number of qubits with $j$th qubit ground state given by $\ket{g_j}$ and excited state by $\ket{e_j}$. The absolute frequencies of these states are given by $\omega_{g_j}$ and $\omega_{e_j}$, respectively. Here $j\in\lbrace 1,N\rbrace$ and $\widetilde{\omega}_{e_{j}}\equiv\omega_{e_{j}}-i\gamma_j/2$, where $\gamma_j$ represents phenomenologically added spontaneous emission rate of the $j$th emitter. The atomic raising $\hat{\sigma}^{\dagger}_j$ and lowering operator $\hat{\sigma}_j$ follow the standard Fermionic anti-commutation relation $\lbrace \hat{\sigma}^{\dagger}_j,\hat{\sigma}_k\rbrace=\delta_{jk}$. With these specifications, $\hat{\mathcal{H}}_{\rm 2LS}$ takes the form
\begin{equation}
\begin{split}
&\hat{\mathcal{H}}_{2LS}=\sum^{N}_{j=1}\big(\hbar\widetilde{\omega}_{e_{j}}\ket{e_j}\bra{e_j}+\hbar\omega_{g_{j}}\ket{g_j}\bra{e_j}\big)\\
&\hspace{9mm}\equiv\hbar\sum^{N}_{j=1}\big(\widetilde{\omega}_{e_{j}}\hat{\sigma}^{\dagger}_j\hat{\sigma}_j+\omega_{g_{j}}\hat{\sigma}_j\hat{\sigma}^{\dagger}_j\big).
\end{split}
\end{equation}
The free-field Hamiltonian $\hat{\mathcal{H}}_{\rm wav}$ incorporates the optical modes of two bi-directional waveguides. In the real-space formalism of quantum optics  \cite{shen2005coherent1}, $\hat{c}_\alpha(x)/\hat{c}^{\dagger}_\alpha(x)$ and $\hat{b}_\alpha(y)/\hat{b}^{\dagger}_\alpha(y)$ (for $\alpha=L$(left),$R$(right)) represent field annihilation/creation operators in the downward and upward waveguides, respectively. These operators follow the usual Bosonic commutation relations $[\hat{c}_\alpha(x),\hat{c}^{\dagger}_\beta(x^{'})]=\delta_{\alpha\beta}\delta(x-x^{'})$ and $[\hat{b}_\alpha(y),\hat{b}^{\dagger}_\beta(y^{'})]=\delta_{\alpha\beta}\delta(y-y^{'})$ where $\beta=L, R$. If we assume group velocity of the field to be $v_{g_d}$ ($v_{g_u}$) in the downward (upward) waveguide then in the linearized dispersion regime $\hat{\mathcal{H}}_{\rm wav}$ can be expressed as
\begin{equation}
\begin{split}
&\hat{\mathcal{H}}_{\rm wav}=-i\hbar v_{g_d}\int dx\Big(\hat{c}^{\dagger}_{R}(x)\partial_x\hat{c}_{R}(x)-\hat{c}^{\dagger}_{L}(x)\partial_x\hat{c}_{L}(x)\Big)\\
&-i\hbar v_{g_u}\int dy\Big(\hat{b}^{\dagger}_{R}(y)\partial_y\hat{b}_{R}(y)-\hat{b}^{\dagger}_{L}(y)\partial_y\hat{b}_{L}(y)\Big).
\end{split}
\end{equation}
$\hat{\mathcal{H}}_{\rm int}$ represents the atom-field interaction Hamiltonian under the rotating wave approximation. Expressing the interaction strength between emitters and upward waveguide (emitters and lower waveguide) by real-valued parameters $V_{\alpha_j}(W_{\alpha_j})$ and specifying emitter locations by the presence of Dirac delta functions on field variables, we write
\begin{equation}
\begin{split}
&\hat{\mathcal{H}}_{\rm int}=\hbar\sum^{N}_{j=1}\sum_{\alpha=L,R}\Bigg[\int dx\delta(x-x_j)\big(V_{\alpha_j}\hat{c}^{\dagger}_\alpha(x)\hat{\sigma}_j+H.c. \big)\\
&+\int dy \delta(y-y_j)\big(W_{\alpha_j}\hat{b}^{\dagger}_{\alpha}(y)\hat{\sigma}_j+H.c. \big)\Bigg].
\end{split}
\end{equation}
Finally, we assume that the separation between the atoms is smaller than the wavelength of the resonant field. This means that in addition to being indirectly coupled through waveguide-mediated interactions, the emitters can also interact directly through infinitely long-ranged dipole-dipole interactions (DDI) \cite{agarwal2012quantum}. With this consideration, the DDI part of the Hamiltonian is given by
\begin{equation}
\hat{\mathcal{H}}_{\rm DDI}=\hbar\sum^{N}_{i=1}\sum\limits_{\substack{j=1 \\ \hspace{-6mm}j > i}}^N J_{ij}\big(\hat{\sigma}^{\dagger}_{i}\hat{\sigma}_j+H.c. \big).
\end{equation}
The strength of the DDI $J_{ij}\equiv J(\mathcal{R}_{ij})$ sensitively depends on the inter-emitter separation $\mathcal{R}_{ij}$ and is expressed as \cite{cheng2017waveguide}
\begin{equation}
\label{DDI}
\begin{split}
&J(\mathcal{R}_{ij})=\frac{3\Gamma_0}{4}\bigg(\frac{\cos \mathcal{R}_{ij}}{\mathcal{R}^3_{ij}}+\frac{\sin \mathcal{R}_{ij}}{\mathcal{R}^2_{ij}}-\frac{\cos \mathcal{R}_{ij}}{\mathcal{R}_{ij}}\bigg)\\
&+\cos^2\theta\bigg(\frac{\cos \mathcal{R}_{ij}}{\mathcal{R}_{ij}}-\frac{3\cos \mathcal{R}_{ij}}{\mathcal{R}^3_{ij}}-\frac{3\sin \mathcal{R}_{ij}}{\mathcal{R}^2_{ij}} \bigg).
\end{split}
\end{equation}
Here $\mathcal{R}_{ij}=\omega_{eg}|\vec{r}_i-\vec{r}_j|/c$, $c$ is the speed of light, $\omega_{eg}=\omega_e-\omega_g$, $\vec{r}_{i/j}$ gives the location of $i/j$th emitter, and $\Gamma_0$ is the free space decay rate of the emitter (also used as the unit in this work). $\theta$ is the angle between the dipole moment $\vec{p}$ of the emitters and the position vector which is defined by $\cos\theta=\vec{p}\cdot(\vec{r}_i-\vec{r}_j)/\lbrace|\vec{p}||\vec{r}_i-\vec{r}_j|\rbrace$.


\section{Photon transport theory}
To investigate the scattering of single photons, the eigenstate $\ket{\Psi}$ of the Hamiltonian ${\bf \hat{\mathscr{H}}}$ in the single excitation manifold of the combined Hilbert space can be constructed as
\begin{equation}
\label{QS}
\begin{split}
&\ket{\Psi}=\Bigg[\sum^{N}_{j=1}\mathcal{A}_j\hat{\sigma}^{\dagger}_j+\sum_{\alpha=L,R}\bigg\lbrace\int dx\varphi_\alpha(x)\hat{c}^{\dagger}_{\alpha}(x)\\
&+\int dy \chi_{\alpha}\hat{b}^{\dagger}_{\alpha}(y)\bigg\rbrace\Bigg]\ket{\varnothing},
\end{split}
\end{equation}
where $\ket{\varnothing}=\ket{g_1,g_2,...,g_N}\otimes \ket{0_{dR},0_{dL}}\otimes\ket{0_{uR},0_{uL}}$ is the ground state of the combined system where all atoms are unexcited and there are zero photons in the upper ($u$) and downward ($d$) waveguides in both left and right directions. $\mathcal{A}_j, \varphi_\alpha$, and $\chi_{\alpha}$ respectively denote the probability amplitudes of finding $j$th emitter in the excited state, photon to be in the bottom and upper waveguide in the $\alpha^{th}$ direction. Initially, we suppose that there no photons in the waveguides, and all QEs are unexcited. For the single-photon routing problem under study, without loss of generality, we assume that the single-photon is launched from the port-(1) and we are interested in port-(1) to port-(3) nonreciprocal routing scheme.

The steady-state analysis of the problem requires the diagonalization of the Hamiltonian $\hat{\mathscr{H}}$. To this end, we insert Eq.~(\ref{Hamil}) and Eq.~(\ref{QS}) in the time-independent Sch\"odinger equation: $\mathscr{\hat{H}}\ket{\Psi}=\hbar\omega\ket{\Psi}$, where $\hbar\omega$ is the energy of the input photon. We obtain the following set of ODEs for the required probability amplitudes
\begin{subequations}\label{eq:AmpEqs}
\begin{align}
-iv_{g_D}\frac{\partial \varphi_{\alpha}(x)}{\partial x}+\sum^{N}_{j=1}V_{\alpha_{j}}\mathcal{A}_{j}\delta(x-x_{j})=\omega\varphi_{\alpha}(x),\\
-iv_{g_D}\frac{\partial \chi_{\alpha}(y)}{\partial y}+\sum^{N}_{j=1}W_{\alpha_{j}}\mathcal{A}_{j}\delta(y-y_{j})=\omega\chi_{\alpha}(y),\\
\sum_{\alpha=L,R}\Big(V_{\alpha_{j}}\varphi_{\alpha}(x_{j})+W_{\alpha_{j}}\chi_{\alpha}(y_{j})\Big)=(\omega-\widetilde{\omega}_{eg_j})\mathcal{A}_{j}\nonumber\\-\sum\limits_{\substack{i=1 \\ \hspace{2mm}i\neq j}}^N J_{ij}\mathcal{A}_i.
\end{align}
\end{subequations}
Here for $\alpha=R, D=d$ and for $\alpha=L, D=u$. In the above equation set, we notice that the DDI interaction appears in the last equation and gives rise to infinitely long-range coupling among all QEs. To solve the Eq. set (\ref{eq:AmpEqs}) we integrate the first equation from $x_k-\epsilon$ to $x_k+\epsilon$ and next equation from $y_k-\epsilon$ to $y_k+\epsilon$. $x_k(y_k)$ represents the location of $k$th emitter for downward (upward) waveguide and $\epsilon << 1$. As a result, we find the following four discontinuity conditions
\begin{subequations}
\label{eq:JumpCond}
\begin{eqnarray}
\varphi_\alpha(x_j+\epsilon)-\varphi_\alpha(x_j-\epsilon)=\frac{-iV_{\alpha_j}}{v_{g_D}}\mathcal{A}_j, \\
\chi_\alpha(y_j+\epsilon)-\chi_\alpha(y_j-\epsilon)=\frac{-iW_{\alpha_j}}{v_{g_D}}\mathcal{A}_j.
\end{eqnarray}
\end{subequations}
Afterward, we establish a connection between $\varphi_\alpha(x)$ and $\varphi_\alpha(x_j\pm\epsilon)$; and similarly between $\chi_\alpha(y)$ and $\chi_\alpha(y_j\pm\epsilon)$ by applying the following regularization relationships
\begin{equation} \label{eq:RegCond}
\begin{split}
&\varphi_\alpha(x_j)=\lim_{\epsilon\rightarrow 0}\bigg[\frac{\varphi_\alpha(x_j+\epsilon)+\varphi_\alpha(x_j-\epsilon)}{2} \bigg], \\
&\chi_\alpha(y_j)=\lim_{\epsilon\rightarrow 0}\bigg[\frac{\chi_\alpha(y_j+\epsilon)+\chi_\alpha(y_j-\epsilon)}{2} \bigg].
\end{split}
\end{equation}

Next, we assume plane wave solutions for the field amplitudes that are modified at each boundary (location of an emitter) by respective transmission, reflection, rightward rectification, and leftward rectification amplitudes. Such an ansatz engenders the following solution 
\begin{equation}
\begin{split}
\label{eq:Ansatz}
&\varphi_R(x)=
\begin{cases}
  t_0e^{iqx}, \hspace{5mm} x<x_1,\\      
  t_{1}e^{iqx}, \hspace{2mm} x_{1} \le x \le x_2, \\
  \vdots \\
  t_{N}e^{iqx}, \hspace{2mm} x>x_{N}  .
\end{cases}\\
&\varphi_L(x)=
\begin{cases}
 r_1e^{-iqx}, \hspace{5mm} x<x_1,\\      
  r_{2}e^{-iqx}, \hspace{2mm} x_{1} \le x \le x_2, \\
  \vdots \\
   r_{N+1}e^{-iqx}, \hspace{2mm} x>x_{N} .
\end{cases}\\
&\chi_R(y)=
\begin{cases}
  \widetilde{t}_0e^{imy}, \hspace{5mm} y<y_1,\\      
  \widetilde{t}_{1}e^{imy}, \hspace{2mm} y_{1} \le y \le y_2, \\
  \vdots \\
  \widetilde{t}_{N}e^{imy}, \hspace{2mm} y>y_{N}  .
\end{cases};\\
&\chi_L(y)=
\begin{cases}
\widetilde{r}_1e^{-imy}, \hspace{5mm} y<y_1,\\      
 \widetilde{r}_{2}e^{-imy}, \hspace{2mm} y_{1} \le y \le y_2, \\
  \vdots \\
  \widetilde{r}_{N+1}e^{-imy}, \hspace{2mm} y>y_{N} .
\end{cases}
\end{split}
\end{equation}
In our model single photons are launched from the port-(1) and there are no other inputs, therefore, we fix $t_0=1, r_{N+1}=0, \widetilde{t}_0=0$ and $\widetilde{r}_{N+1}=0$. For a chain of identical emitters the optical  wavenumbers $q$ and $m$ are defined through $q=(\omega-\omega_{eg})/v_{g_d}$ and $m=(\omega-\omega_{eg})/v_{g_u}$, respectively. The relation among various amplitudes can be found by inserting Eq.~(\ref{eq:Ansatz}) into Eq.~(\ref{eq:JumpCond}) and Eq.~(\ref{eq:AmpEqs}) which after simplifications produce the following set of recurrence relations
\begin{subequations}\label{eq:RecRel}
\begin{align}
t_j-t_{j-1}=\frac{-iV_{R_j}}{v_{g_d}}\mathcal{A}_j e^{-iqx_j}, \\
r_{j+1}-r_{j}=\frac{iV_{L_j}}{v_{g_d}}\mathcal{A}_j e^{iqx_j}, \\
\widetilde{t}_j-\widetilde{t}_{j-1}=\frac{-iW_{R_j}}{v_{g_u}}\mathcal{A}_j e^{-imy_j}, \\
\widetilde{r}_{j+1}-\widetilde{r}_{j}=\frac{iW_{L_j}}{v_{g_u}}\mathcal{A}_j e^{imy_j}, \\
\frac{V_{R_j}}{2}e^{iqx_j}\big(t_j+t_{j-1}\big)+\frac{V_{L_j}}{2}e^{-iqx_j}\big(r_{j+1}+r_{j}\big)\nonumber\\
+\frac{W_{R_j}}{2}e^{imy_j}\big(\widetilde{t}_j+\widetilde{t}_{j-1}\big)+\frac{W_{L_j}}{2}e^{-imy_j}\big(\widetilde{r}_{j+1}+\widetilde{r}_{j}\big)\nonumber\\
=(\omega-\widetilde{\omega}_{eg_j})\mathcal{A}_{j}-\sum\limits_{\substack{i=1 \\ \hspace{2mm}i\neq j}}^N J_{ij}\mathcal{A}_i.
\end{align}
\end{subequations}
The required net transmission, reflection, rightward rectification and leftward rectification intensities (which can be calculated by solving the above set of coupled equations for any $j$ value) are defined by $T= |t_N e^{iqNL}|^{2}$, $R=|r_1|^{2}$, $\widetilde{T}=|\widetilde{t}_N e^{imNL}|^{2}$, and $\widetilde{R}=|\widetilde{r}_1|^{2}$, respectively. It is worthwhile to mention that the time-delays (retardation) effects are already incorporated in our model due to the presence of propagation phases $e^{iqx_j}$ and $e^{imy_j}$. For chiral cases, these phases don't appear in the output intensities because of the unidirectionality of the problem. However, when the direct back reflections between QEs are added due to DDI then these retardation phases impact the photon transport.


\section{Results}
There are various experimental platforms where the results of our generic theoretical model can be applied. Examples include superconducting Josephson junctions (artificial atoms) in microwave transmission lines \cite{schoelkopf2008wiring}, Cesium atoms coupled to photonic crystal waveguides \cite{hood2016atom}, and semiconducting quantum dots (QDs) interacting with nanowires (nanofibers) \cite{akimov2007generation}. Following experimental setups studied in Ref.~\cite{akimov2007generation, chang2006quantum, cheng2017waveguide}, we consider a chain of QDs interacting with two identical $Ag$ nanowires. The transition wavelength $\lambda_{QD}$ is taken to be $655nm$ with spontaneous emission rate $\gamma=6.86\Gamma_0$ and emitter-waveguide coupling strength $\Gamma=11.03\Gamma_0$. The free space decay rate $\Gamma_0$ is allotted a value of $7.5 MHz$. 

\begin{figure*}
\centering
  \begin{tabular}{@{}cccc@{}}
   \includegraphics[width=3in, height=2in]{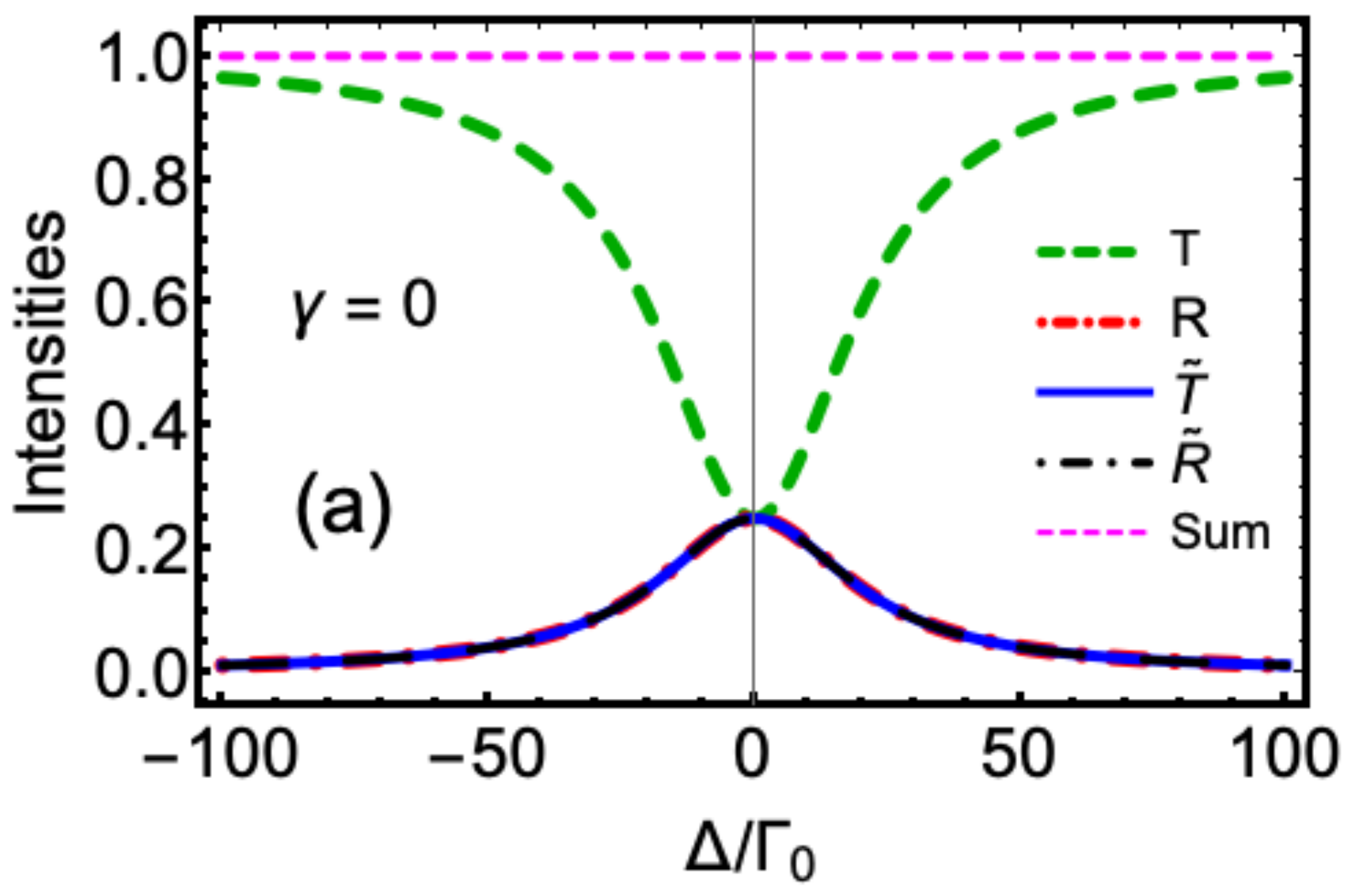} &
   \includegraphics[width=3in, height=2in]{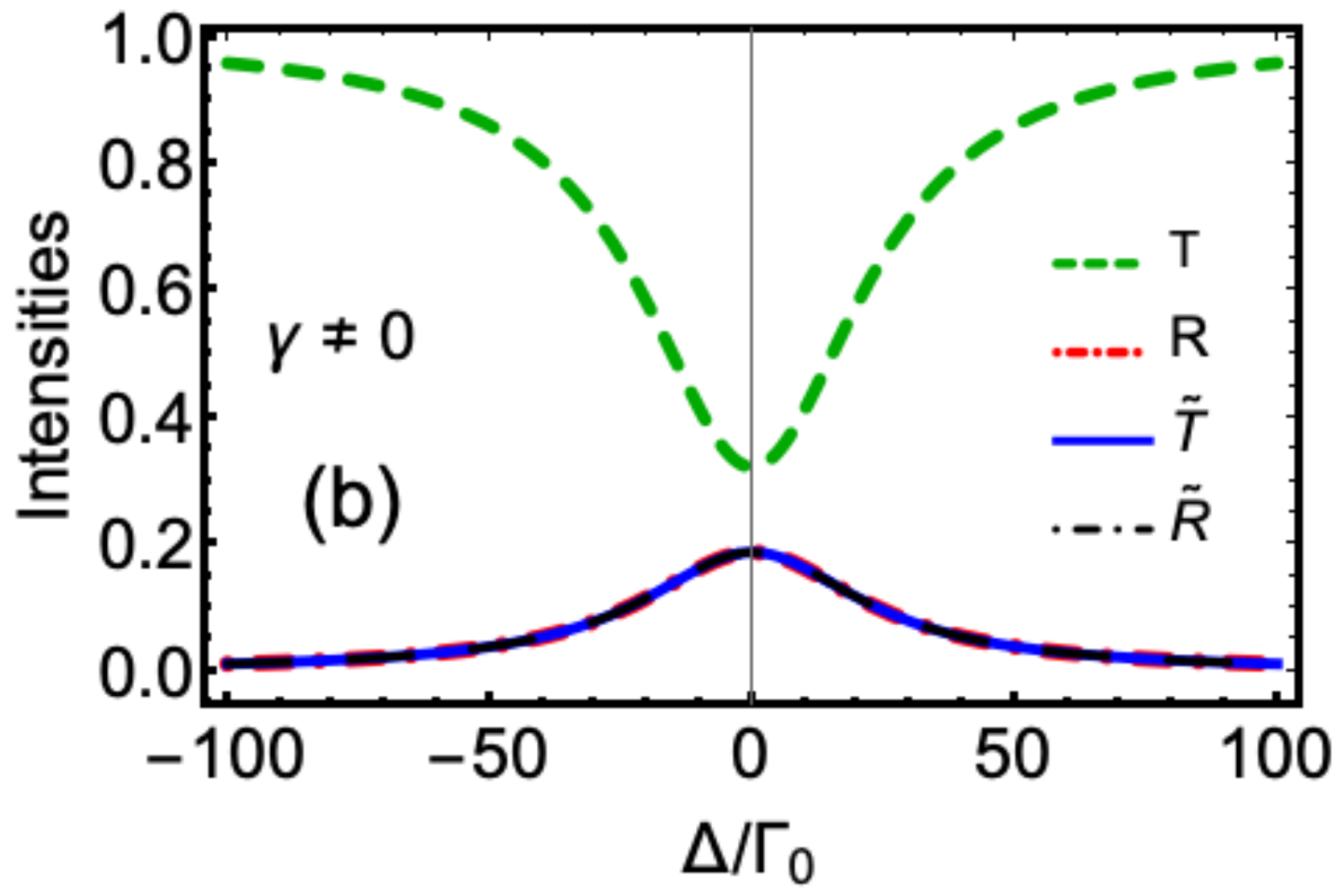} 
  \end{tabular}
\captionsetup{
  format=plain,
  margin=0.1em,
  justification=raggedright,
  singlelinecheck=false
}
 \caption{(Color online) Transport intensities versus detuning under symmetric coupling scenario. For plot (a) we ignore the spontaneous emission i.e. $\gamma=0$, while in (b) $\gamma\neq 0$. The emitter-waveguide coupling strength is fixed to $\Gamma=11.03\Gamma_0$ while spontaneous emission rate $\gamma$ is $6.86\Gamma_0$.}\label{Fig2}
\end{figure*}

The inter-emitter separation appears at two places in our calculations: (I) In the expression of DDI (II) In the free propagation phases ($qx_j$ and $my_j$) accounting for the retardation effects. In the DDI expression, for the sake of simplicity, we assume the dipole moment of all QDs to be perpendicular to the position vector direction implying $\cos \theta=0$. For any two consecutive emitters at positions $(0, 17nm, 0)$ and $(32.75nm, 17nm, 0)$ (lattice constant of $32.75nm$), from Eq.~(\ref{DDI}) the DDI interaction turns out to be $J_{12}=J_{21}\equiv J=23.10\Gamma_0$. For the free propagation phase, we consider the wavelength of the surface plasmon $\lambda_{sp}$ to be $211.8nm < \lambda_{QD}$ due to a reduction in the group velocity in the waveguide. If we assume both waveguides to be identical then these phases take the value $qx_j=my_j\equiv\Theta=0.31\pi$.

\subsection{Routing with a single emitter}
Let us begin with the simplest case of a single emitter. In the following, we focus on two scenarios, namely, symmetric coupling and chiral coupling cases. 
\subsubsection{Symmetric coupling case}
In the symmetric case, we take the same value of emitter-waveguide coupling towards left and right i.e. $V_R=V_L=W_R=W_L\equiv \mathcal{U}$ and $v_{g_d}=v_{g_u}\equiv v_g$ such that $\Gamma=\mathcal{U}^2/v_g$. The parameter $\Gamma$ is defined to quantify emitter-waveguide coupling strength. One can find the transport amplitudes by inserting $j=1$ in Eq.~(\ref{eq:RecRel}) and solve for $t_1\equiv t,r_1\equiv r,\widetilde{t}_1\equiv\widetilde{t},$ and $\widetilde{r}_1\equiv\widetilde{r}$. We obtain
\begin{subequations}\label{Tsym1}
\begin{align}
t=\frac{-8\Gamma^3-4\Gamma^2(\gamma-2i\Delta)+2\Gamma(\gamma-2i\Delta)^2+(\gamma-2i\Delta)^3}{\big(\gamma-2\Gamma-2i\Delta\big)(\gamma+2\Gamma-2i\Delta)(\gamma+4\Gamma-2i\Delta)},\\
r=\widetilde{t}=\widetilde{r}=\bigg(\frac{-2\Gamma(2\gamma-2i\Delta)+4\Delta^2-\gamma^2+4i(\gamma+\Gamma)\Delta}{2\Gamma(\gamma-2i\Delta)} \bigg)^{-1}.
\end{align}
\end{subequations}
In Fig.~\ref{Fig2}(a), we plot probabilities of detecting photon at different output ports as a function of detuning $\Delta\equiv\omega-\omega_{eg}$. For this plot, we set $\gamma=0$. As expected we find $T+R+\widetilde{T}+\widetilde{R}=1$ ensuring proper normalization. More importantly, from Eq.~(\ref{Tsym1}) we notice that at $\Delta=0$ all intensities share a common value of $0.25$ which indicates an equal probability of detecting photon at all four ports (as also reported in \cite{zhou2013quantum, yan2018targeted}). Such behavior is completely contrary to the required routing protocol where the photon is expected to emerge at port-(3) with $100\%$ probability. Under far-off resonance conditions, we find that the transmission into port-(2) reaches almost 100\% at the expense of decreasing intensities at other ports. A complete mismatch between incoming photon frequency and $\omega_{eg}$ results in such behavior where photon continue to propagate in the bottom waveguide where eventually it is detected at port-(2) without being routed.

In Fig.~\ref{Fig2}(b), we introduce a finite spontaneous emission rate. With $\gamma=6.86\Gamma_0$, we notice that the on-resonance output intensities $T=\widetilde{T}=\widetilde{R}$ decreases to $\sim19\%$ while reflection intensity from the bottom waveguide $\widetilde{R}$ raises to $\sim32\%$. The remaining probability contribution is lost to non-waveguide (environmental) modes. From Fig.~\ref{Fig2}(a) and Fig.~\ref{Fig2}(b) we conclude that symmetric coupling fails to attain deterministic routing. But what will happen if we could break the symmetry in QE emission into the waveguide modes? Next, we answer this question by taking advantage of the chiral photon emissions.

\subsubsection{Chiral coupling case}
For the chiral case, we suppose that the QE is allowed to interact only with the right propagating modes in both waveguides with back-reflection channels completely blocked, i.e. we set $\Gamma_{dL}\equiv V^{2}_{L}/v_{g_d}=0$, $\Gamma_{uL}\equiv W^{2}_{L}/v_{g_u}=0$ with $\Gamma_{dR}=\Gamma_{uR}=\Gamma$ while $\Gamma_{dR}\equiv V^{2}_{R}/v_{g_d}$ and $\Gamma_{uR}\equiv W^{2}_{R}/v_{g_u}$. 
\begin{figure*}
\centering
  \begin{tabular}{@{}cccc@{}}
   \includegraphics[width=3in, height=2in]{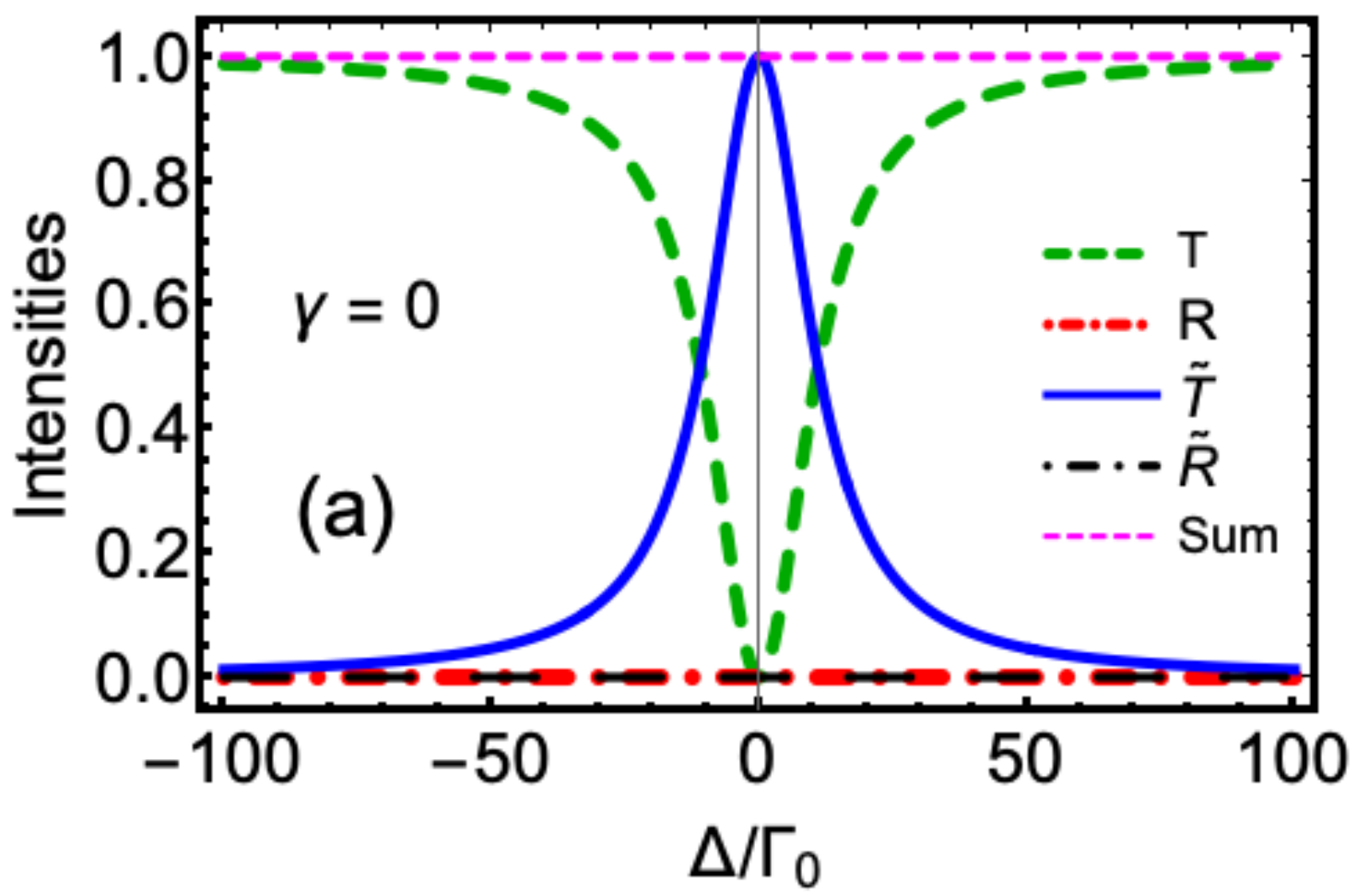} &
  \hspace{4mm}\includegraphics[width=3in, height=2in]{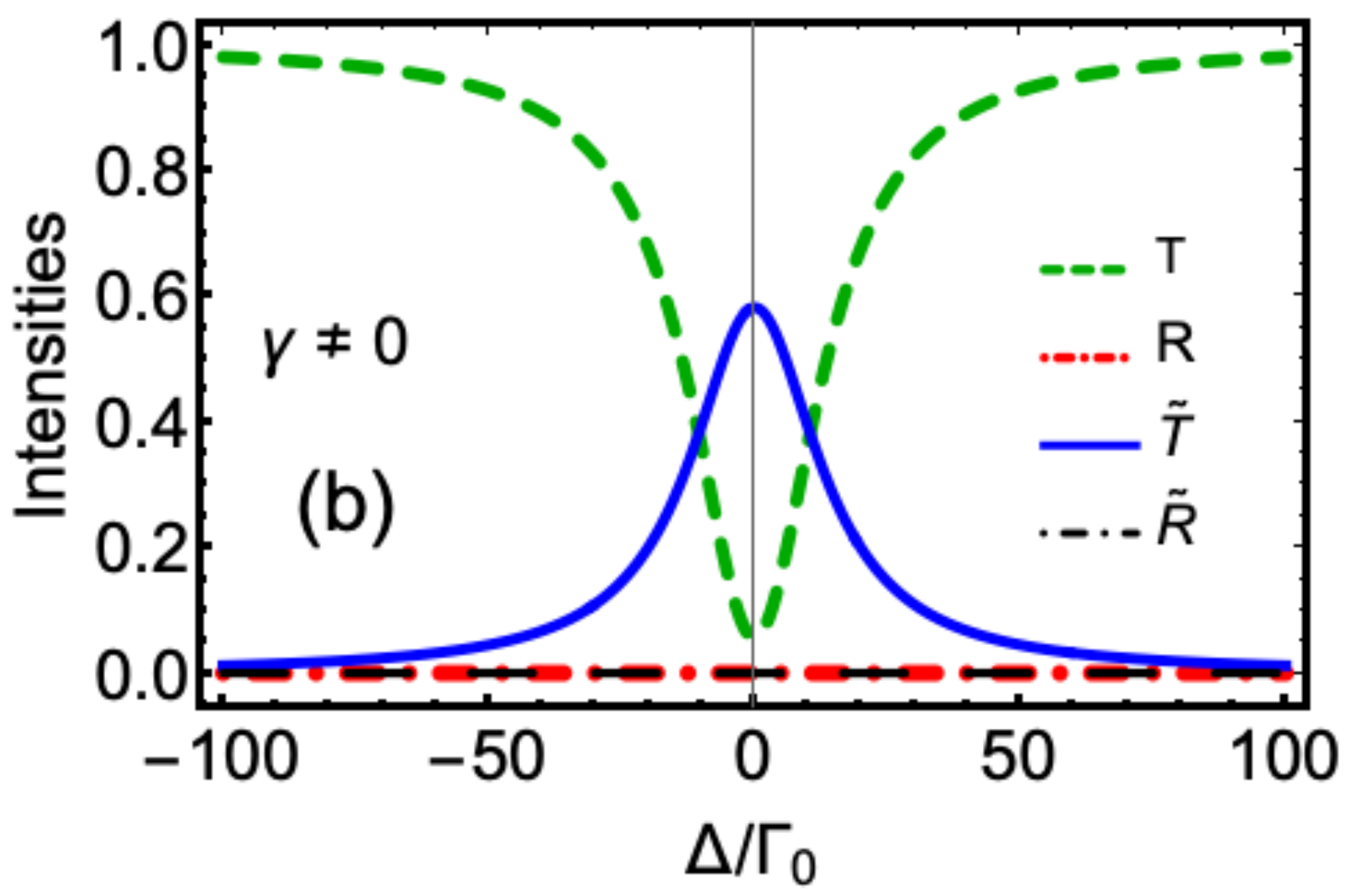} 
  \end{tabular}
\captionsetup{
  format=plain,
  margin=0.1em,
  justification=raggedright,
  singlelinecheck=false
}
 \caption{(Color online) Port intensities as a function of detuning for a perfect chiral situation in which photon emission in the left direction is not allowed (a) $\gamma=0$ (b) $\gamma\neq0$. In this plot $\Gamma=11.03\Gamma_0$ and $\gamma=6.86\Gamma_0$.}\label{Fig3}
\end{figure*}
Under these conditions, the transport amplitudes take the form
\begin{equation}\label{Tchr1}
t=\bigg(\frac{\gamma-2i\Delta}{\gamma+2\Gamma-2i\Delta}\bigg),
r=0;
\widetilde{t}=\bigg(\frac{-2\Gamma}{\gamma+2\Gamma-2i\Delta}\bigg),
\widetilde{r}=0.
\end{equation}
In Fig.~\ref{Fig3}, we plot the output intensities for such a chiral setting. Fig.~\ref{Fig3}(a) concentrates on the no loss situation. We note that at $\Delta=0$ the transmission from the bottom waveguide $T$ vanishes. As pointed out in references \cite{shen2005coherent1, gonzalez2016nonreciprocal}, such behavior is attributed to the perfect destructive interference between the amplitudes of incoming photons and the photons emitted into the forward direction of the bottom waveguide. Consequently, the rightward rectification $\widetilde{T}$ takes a unit value with the photon emerging at the port-(3) with acquiring a phase shift of $\pi$. Thus we conclude that chiral emitter-waveguides couplings can be used for perfect photo-routing if the conditions of on-resonance and $\gamma=0$ are met. 

However, in any realistic experimental proposal spontaneous emission cannot be ignored. With this concern, in Fig.~\ref{Fig3}(b) we present intensities with $\gamma\neq 0$. Form Eq.~(\ref{Tchr1}), we find the transport amplitudes take the values $t=(1+2\Gamma/\gamma)^{-1}$ and $\widetilde{t}=-(1+0.5 \gamma/\Gamma)^{-1}$ at resonance. We notice with a $\gamma$ value of about $62\%$ of the emitter-waveguide coupling rate, even in the presence of chirality and perfect resonance conditions, the efficiency of the routing scheme drastically reduces to $58\%$. Additionally, $T$ raises to $\sim 6\%$. From this result, it is evident that the deterministic nature of the routing scheme achieved by chirality is lost when QEs are allowed to spontaneously emit the photons.

\subsection{Routing with collective response of two QEs}
\begin{figure*}
\centering
  \begin{tabular}{@{}cccc@{}}
    \includegraphics[width=3.1in, height=1.6in]{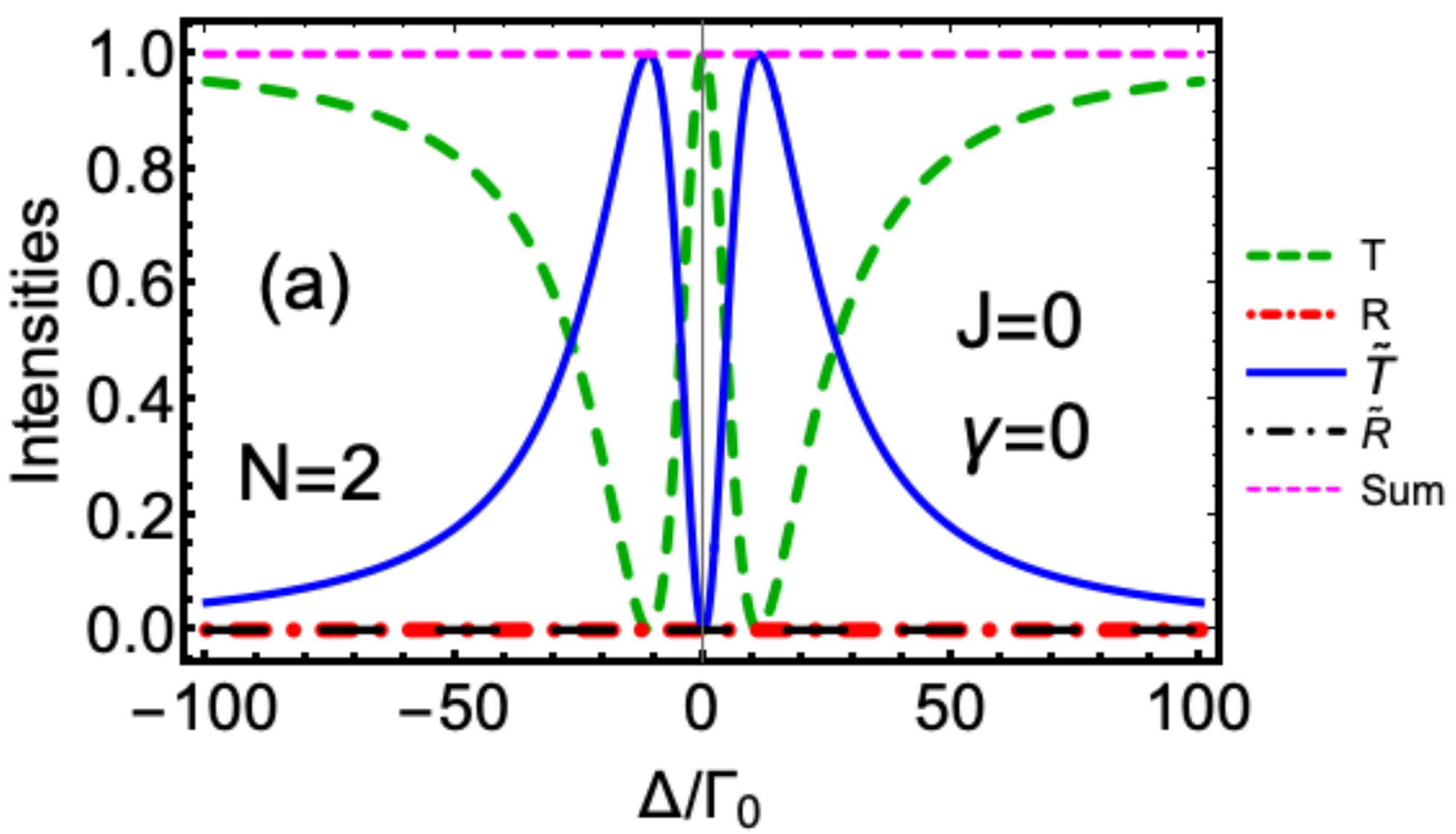} 
   \includegraphics[width=2.5in, height=1.6in]{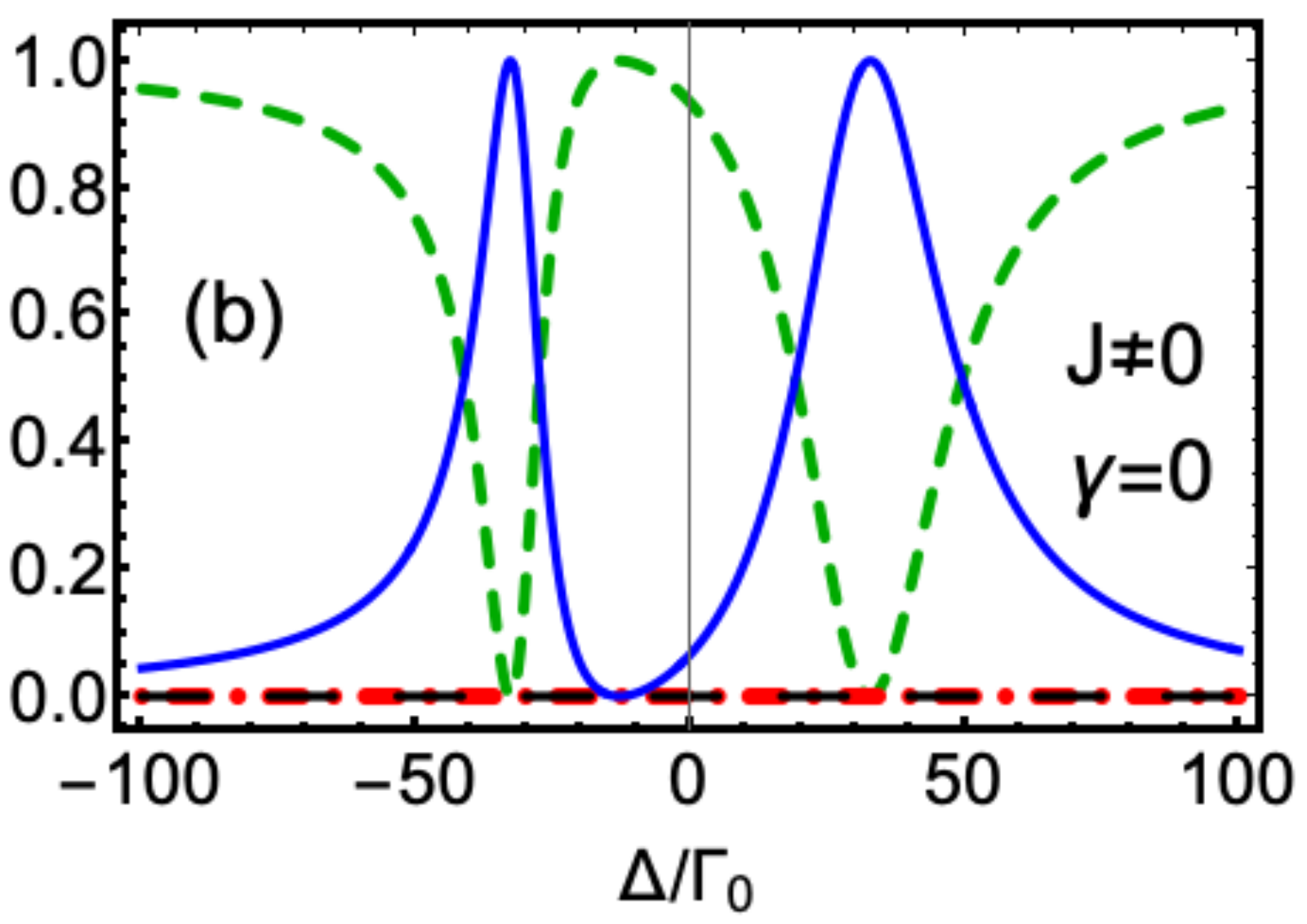}\\
   \includegraphics[width=2.60in, height=1.7in]{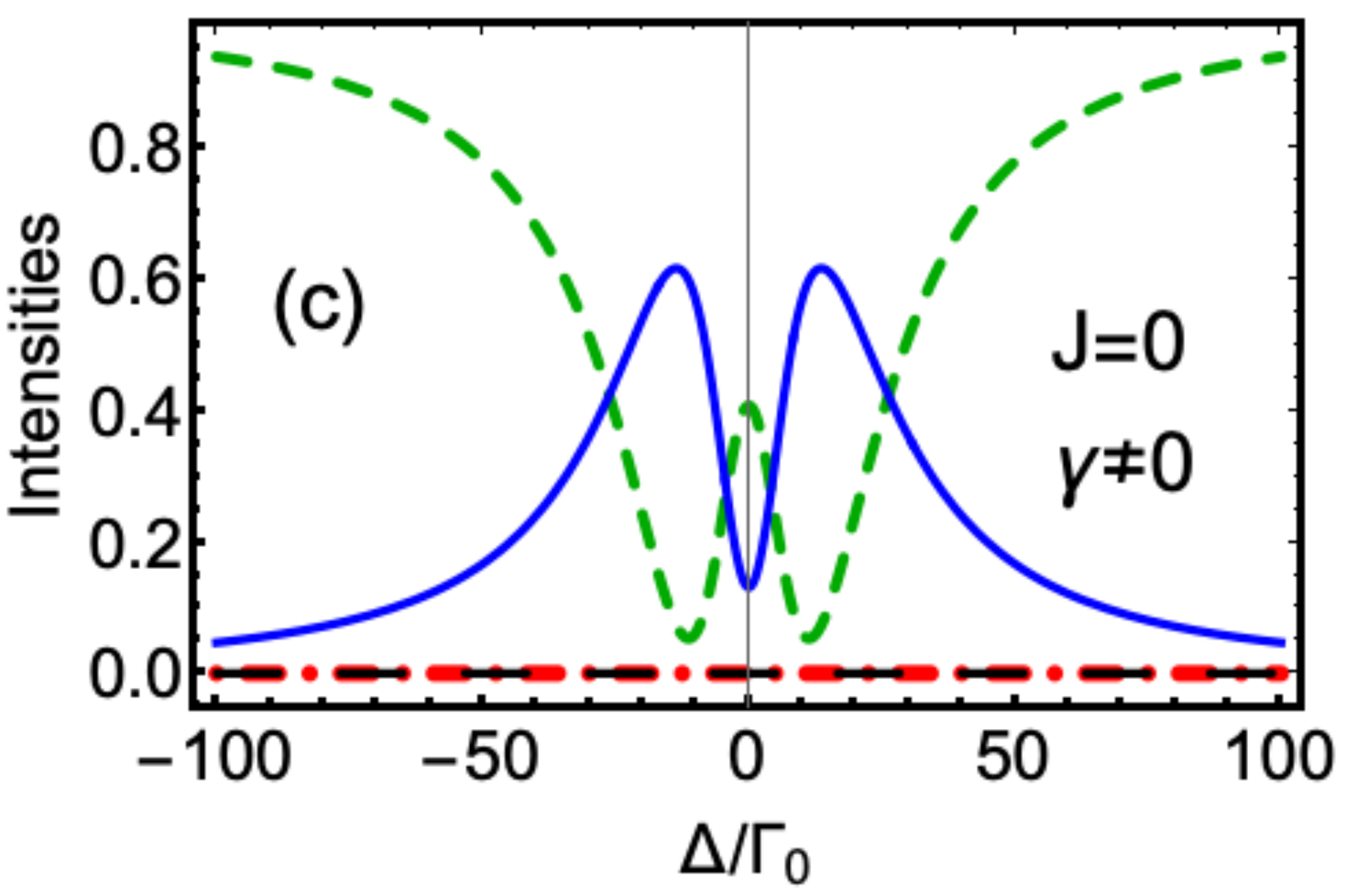} 
    \hspace{12mm}\includegraphics[width=2.5in, height=1.7in]{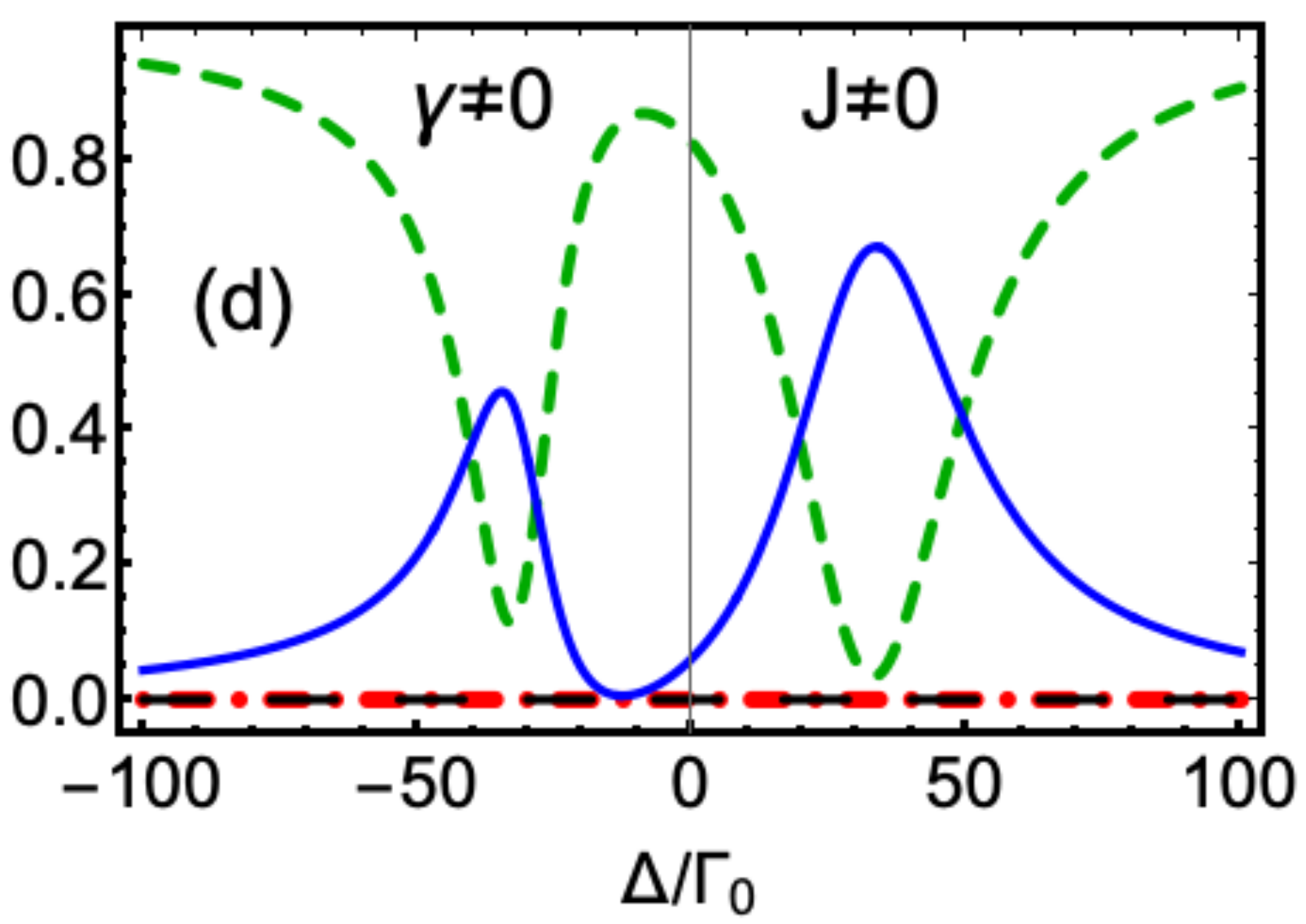} 
  \end{tabular}
\captionsetup{
  format=plain,
  margin=0.1em,
  justification=raggedright,
  singlelinecheck=false
}
 \caption{(Color online) Single-photon $T$, $\widetilde{T}$, $R$, and $\widetilde{R}$ spectra/intensities for two QEs separated by distance $mL=qL=\pi$ as a function of detuning $\Delta$ for a chiral coupling. Four possible cases of DDI between QEs ($J$) and spontaneous emission rate ($\gamma$) are shown. In (a) $J=0$ and $\gamma=0$ (b) $\gamma=0$ and  emitters are $\lambda_{sp}/20$ apart which, according to Eq.(\ref{DDI}), gives rise to $J=23.10\Gamma_0$ (c) $J=0$ and $\gamma=6.86 \Gamma_0$ (d) $J=23.10 \Gamma_0$ and $\gamma=6.86 \Gamma_0$. In all plots $\Gamma=11.03\Gamma_{0}$.}\label{Fig4}
\end{figure*}

We now concentrate on the question, which is central to this work, that in what ways the routing of photons can be improved in the presence of dissipative loss from QEs chirally coupled to a waveguide ladder? To this end, we first analyze the routing in the presence of two QEs simultaneously interacting through waveguide modes and DDI. From Eq.~(\ref{eq:RecRel}) we obtain the analytic expressions for net transmission $t_2$ and rightward rectification amplitude $\widetilde{t}_2$ as
\begin{subequations}\label{2QEs}
\begin{align}
t_2=\frac{-2i\sin\Theta J\Gamma+i\big(4J^2+\gamma^2+4\Gamma^2-4i\gamma\Delta-4\Delta^2\big)}{4iJ^2+8e^{i\Theta}J\Gamma+i\big(\gamma+2\Gamma-2i\Delta\big)^2},\\
\widetilde{t}_2=\frac{-4e^{-i\Theta}J\Gamma\big(1+e^{2i\Theta}\big)-4\Gamma\big(i\gamma+2\Delta\big)}{4iJ^2+8e^{i\Theta}J\Gamma+i\big(\gamma+2\Gamma-2i\Delta\big)^2}.
\end{align}
\end{subequations}
In Fig.~\ref{Fig4} we present the probabilities of photon detection at all four ports. In these plots, we mainly concentrate on two parameters, DDI ($J$) and spontaneous emission rate ($\gamma$). In Fig.~\ref{Fig4}(a) we present the simplest case when both QEs are not coupled through DDI and spontaneous emission has been disregarded. For this situation, Eq.~(\ref{2QEs}) simplifies to
\begin{equation}\label{2QEsJZGZ}
t_2=\frac{\Gamma^2-\Delta^2}{\big(\Gamma-i\Delta\big)^2},\hspace{15mm}
\widetilde{t}_2=\frac{2i\Gamma\Delta}{\big(\Gamma-i\Delta\big)^2}.
\end{equation}
We note that the term $\Gamma$ appearing in the demonstrator of $t_2$ and $\widetilde{t}_2$ expressions in Eq.~(\ref{2QEsJZGZ}) represent an effective DDI between QEs mediated through the waveguides (see also \cite{cheng2017waveguide} for a similar argument). And this effective DDI influences the spectral lineshape and gives rise to a splitting of peaks in Fig.~(\ref{Fig4}a). We further notice, at resonance, there is no rightward rectification but the system shows complete transparency from bottom waveguide i.e. $t$ reaches a unit value. On the other hand, it follows from Eq.~(\ref{2QEsJZGZ}) as well as from Fig.~\ref{Fig4}(a), that when $\Delta=\pm\Gamma$ the bottom waveguide transmission vanishes, while rightward rectification takes a unit value. Combined, these features generate a frequency separation between both peaks of value $2\Gamma$. It is worthwhile to point out that even without DDI, two QEs can achieve deterministic routing at two different (symmetric) values of detuning which was not possible otherwise with a single QE. 

In the next case, we still neglect the spontaneous emission but consider a finite DDI between the QEs. For this situation Eq.~(\ref{2QEs}) gives
\begin{equation}\label{2QEsJZ}
\begin{split}
&t_2=\frac{-2\Gamma \big(\cos\Theta J+\Delta\big)}{iJ^{2}+2e^{i\theta}J\Gamma+i\big(\Gamma-i\Delta\big)},\\
&\widetilde{t}_2=\frac{2iJ\Gamma\sin\Theta+i(J^2+\Gamma^2-\Delta^2)}{iJ^{2}+2e^{i\theta}J\Gamma+i\big(\Gamma-i\Delta\big)}.
\end{split}
\end{equation}
In the above amplitudes, we notice that the presence of DDI results in a non-vanishing value of free propagation phase $\Theta$ which vanished for the $J=0$ case where the photon propagates in a unidirectional/chiral fashion in the waveguides \cite{mirza2017chirality, mirza2013single}. Fig.~\ref{Fig4}(b) shows the spectra for this case. We notice that the DDI lifts the symmetry in peak locations of $\widetilde{T}=\vert \widetilde{t}_2\vert^2$ with a broadened peak on the positive $\Delta$ axis. Rightward rectification approaches a null value at $\Delta=-cos\Theta J$ while similar to Fig.~\ref{Fig4}(a) there are two DDI-dependent frequencies where perfect routing is achievable. These frequencies appear at $\Delta=\pm\sqrt{J^2+\Gamma^2+2J\Gamma\sin\Theta}$ which are two roots of maximized $\widetilde{T}$. This behavior suggests that by altering the QE separation we can tune the frequency values where perfect routing is attained.

Subsequently, we take a more realistic scenario in which QEs are allowed to dissipate through spontaneous emission but there is no DDI i.e. $J=0$ (see Fig~\ref{Fig4}(c)). For this particular case, Eq.~(\ref{2QEs}) yields
\begin{equation}\label{2QEsJZ}
t_2=\frac{\gamma^2+4\Gamma^2-4i\gamma\Delta-4\Delta^2}{\big(\gamma+2\Gamma-2i\Delta\big)^2},\hspace{2mm}
\widetilde{t}_2=\frac{-4\Gamma\big(\gamma-2i\Delta\big)}{\big(\gamma+2\Gamma-2i\Delta\big)^2}.
\end{equation}
As expected, with a non-zero $\gamma$ value, $\widetilde{t}_2$ reduces drastically and reaches to almost $62\%$. Spontaneous emission also contributes to the widths of the peaks and shifts the peak values slightly away from their previous value of $\pm\Gamma$ (as shown in Fig.~(\ref{Fig4}a)). Our parameter choice assumes an over-coupled regime i.e. $\Gamma>\gamma$. Under this condition, one can develop a partial understanding of intensities analytically. For a small dissipation i.e. $\Gamma>\gamma$, at resonance, the approximate form of intensities is given by $t_2\approx 4\Gamma^2/4\Gamma^2\rightarrow 1$ and $\widetilde{t}_2\approx-\gamma/\Gamma\rightarrow0$. We observe a similar tendency of intensities in Fig.~(\ref{Fig4}c) near $\Delta=0$.

\begin{figure*}
\centering
  \begin{tabular}{@{}cccc@{}}
   \includegraphics[width=3.3in, height=2.3in]{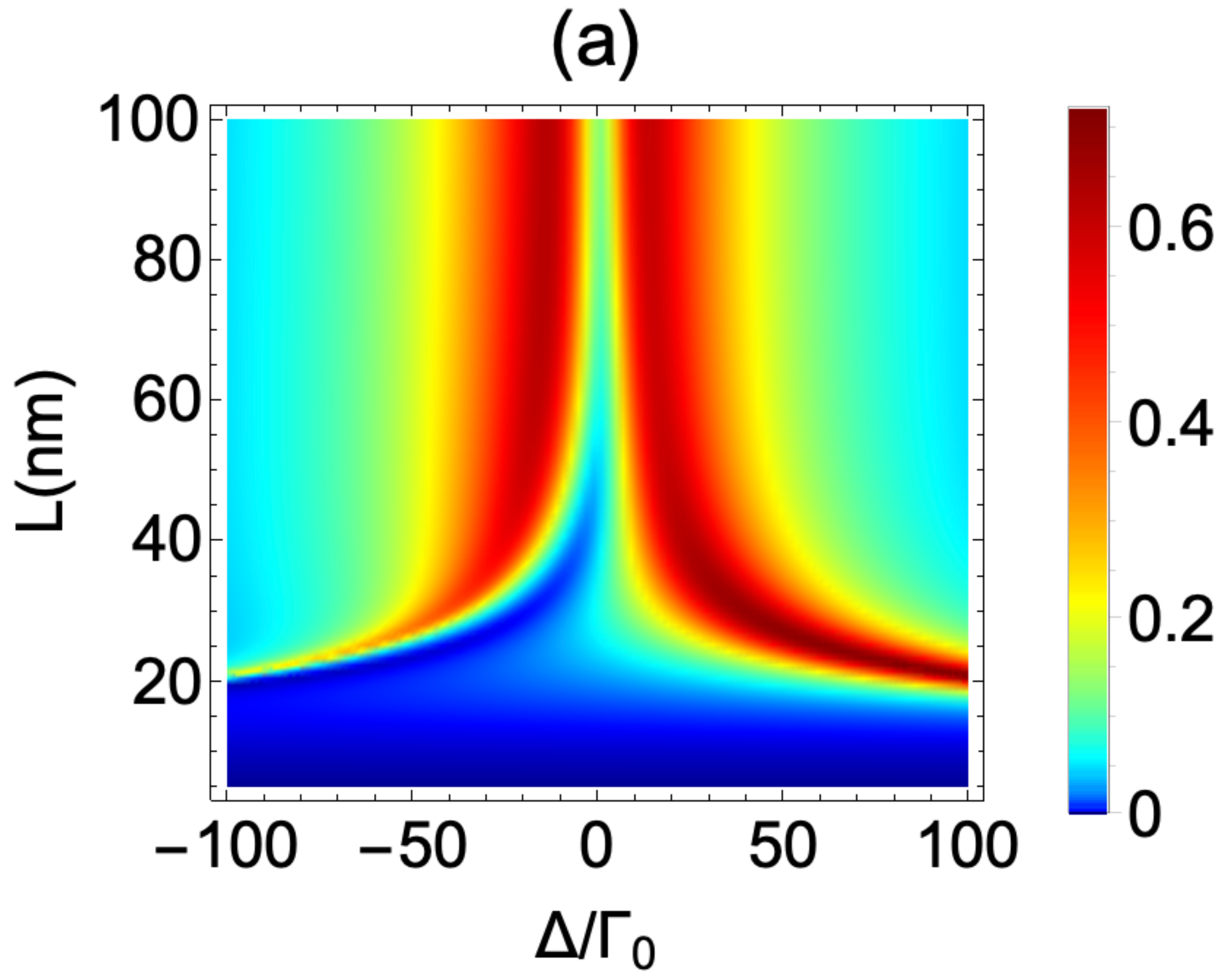} &
  \hspace{2mm}\includegraphics[width=3.2in, height=2.32in]{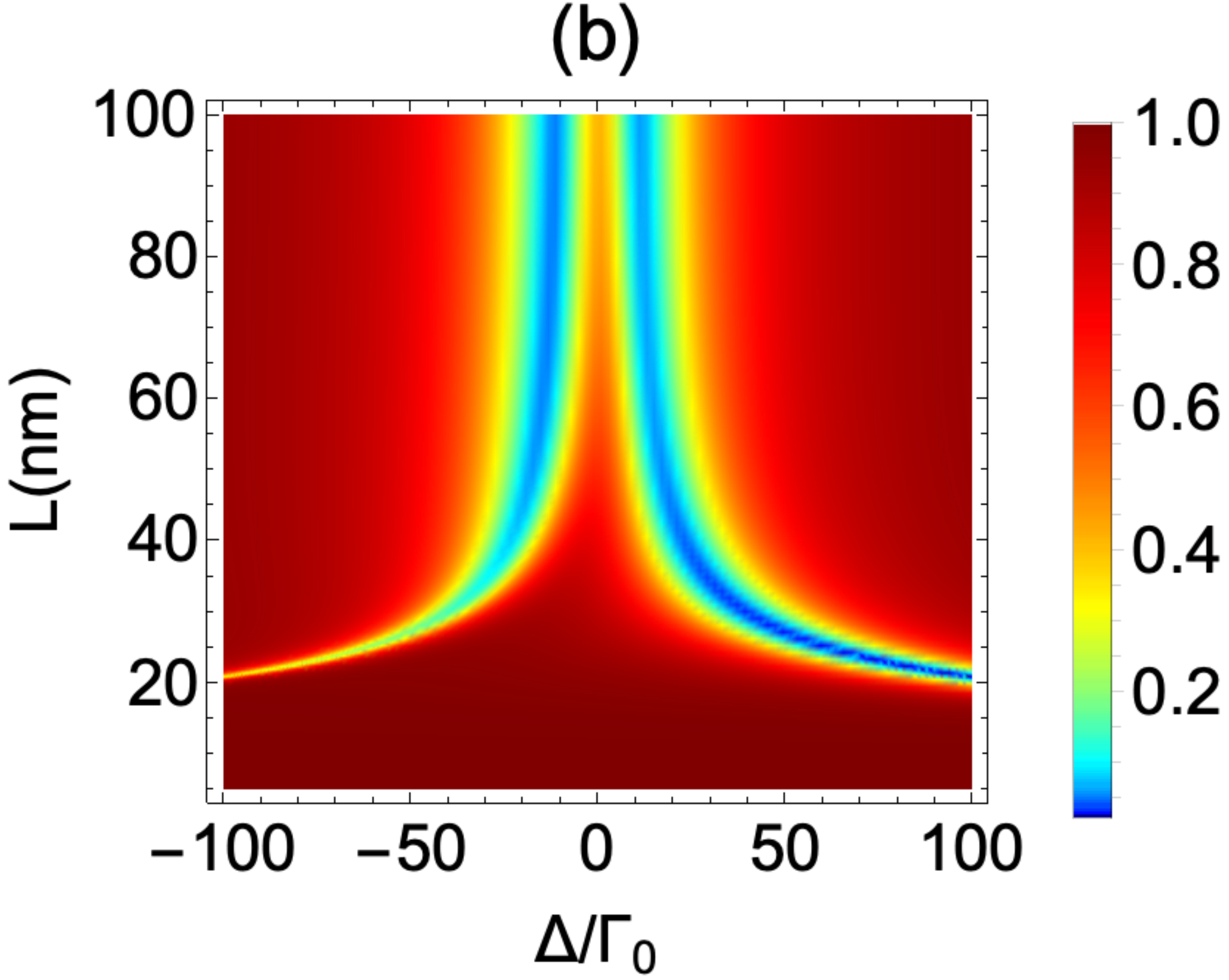} 
  \end{tabular}
\captionsetup{
  format=plain,
  margin=0.1em,
  justification=raggedright,
  singlelinecheck=false
}
 \caption{(Color online) Density plot showing the single-photon (a) rightward rectification ($\widetilde{T}$) and (b) lower waveguide transmission ($T$) as a function of detuning $\Delta$ and inter-emitter separation $L$ for a system of $2$ QEs chirally coupled to a waveguide ladder. The emitter-waveguide coupling is chosen to be $\Gamma=11.03\Gamma_0$ with spontaneous emission rate $\gamma=6.86\Gamma_0$. }\label{Fig5}
\end{figure*}

Finally, in Fig.~(\ref{Fig4}d) we present the intensities when both $J$ and $\gamma$ are non-zero. We notice that due to spontaneous emission symmetry in the peak heights is lost (see Fig.~(\ref{Fig4}b) for comparison). This feature, as also pointed out in Ref.~\cite{cheng2017waveguide}, is attributed to the energy loss introduced by the atomic dissipation. However, when we compare Fig.~(\ref{Fig4}c) and Fig.~(\ref{Fig4}d) from the routing perspective, we notice that the rightward rectification reaches 67\% which is 5\% better than the no DDI case and 9\% better than the single-emitter case. Albeit, the peak location for the maximum redirection of the photon from bottom to upper waveguide is shifted towards positive detuning. This result clearly indicates that even for $\gamma\neq 0$, DDI opens the possibility of routing efficiency enhancement at one of the spectral peaks.

In Fig.~\ref{Fig4} we considered a single value of DDI i.e. $J=23.10\Gamma_0$ due to inter-emitter separation choice of $L=32.75nm$. To see how DDI can influence the routing as a function of $\Delta$, in Fig~\ref{Fig5} we plot port intensities for a range of DDI by varying $L$ value between $5nm$ to $100nm$. We notice that the DDI considerably changes $\widetilde{T}$ and consequently $T$. In particular, above $L\sim 20nm$ we find two regions between $-40\Gamma_0\lesssim \Delta \lesssim 0$ and $0\lesssim \Delta \lesssim 40\Gamma_0$ where the probability of photon detection at port-(3) is at least $60\%$ i.e. $\widetilde{T}\geq60\%$. Correspondingly in the same regions, lower waveguide transmission $T$ remains less than $20\%$ approximately. Outside these regions the behavior of $\widetilde{T}$ and $T$ shows opposite trends, indicating a progressive deterioration in the photon routing.

\begin{figure*}
\centering
  \begin{tabular}{@{}cccc@{}}
  \includegraphics[width=3.1in, height=1.6in]{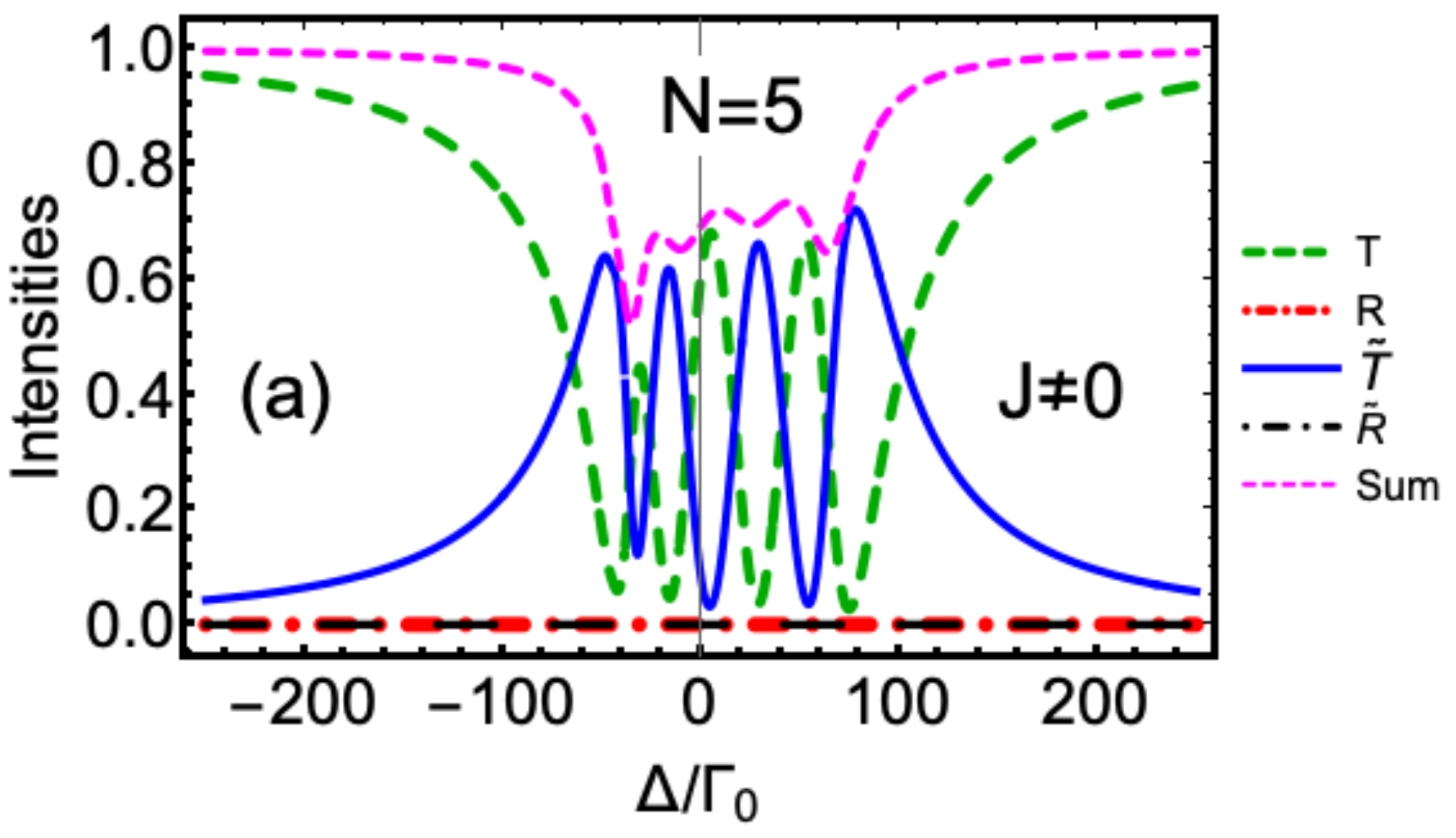} 
   \includegraphics[width=2.5in, height=1.6in]{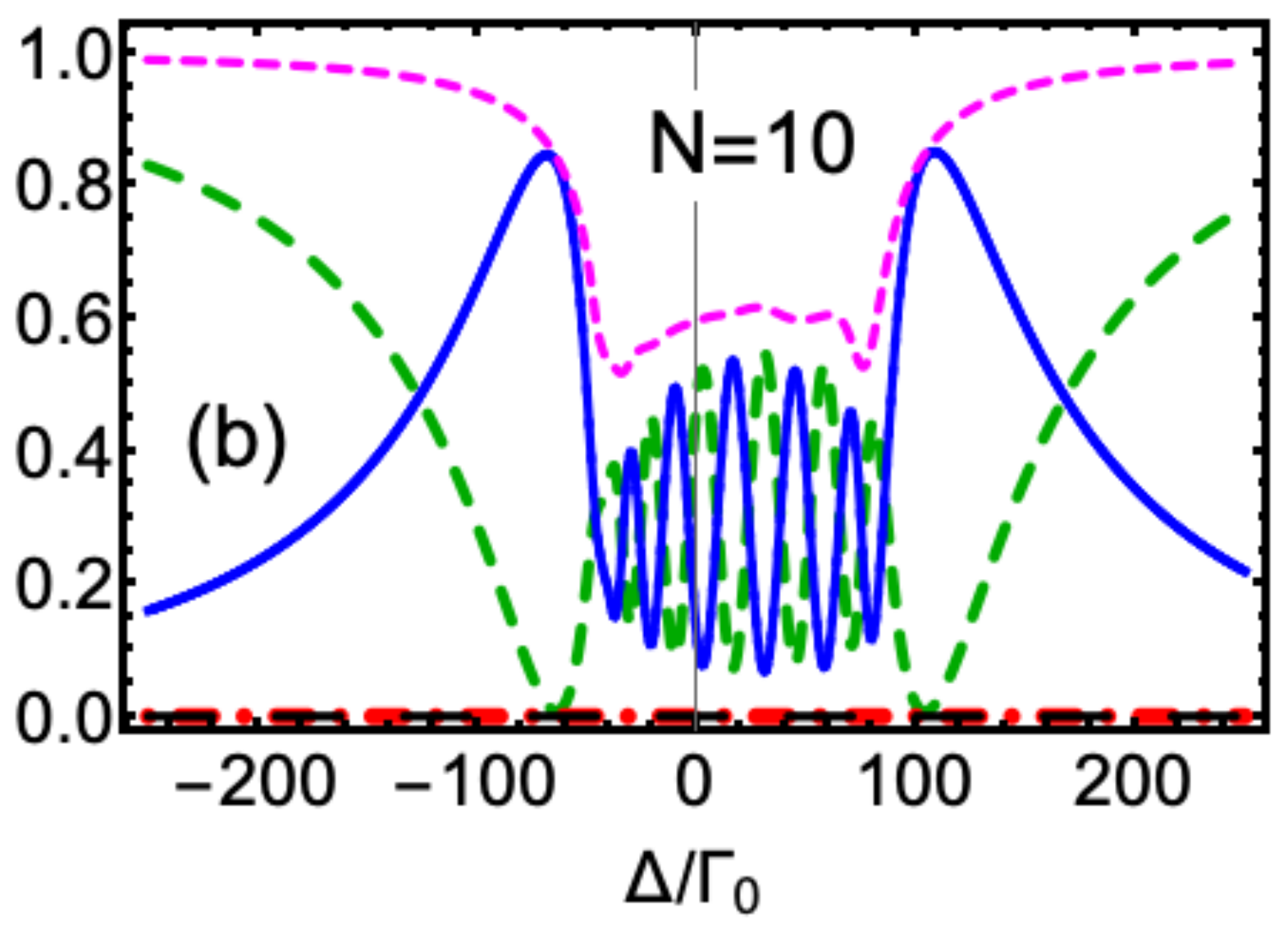}\\
   \includegraphics[width=2.60in, height=1.7in]{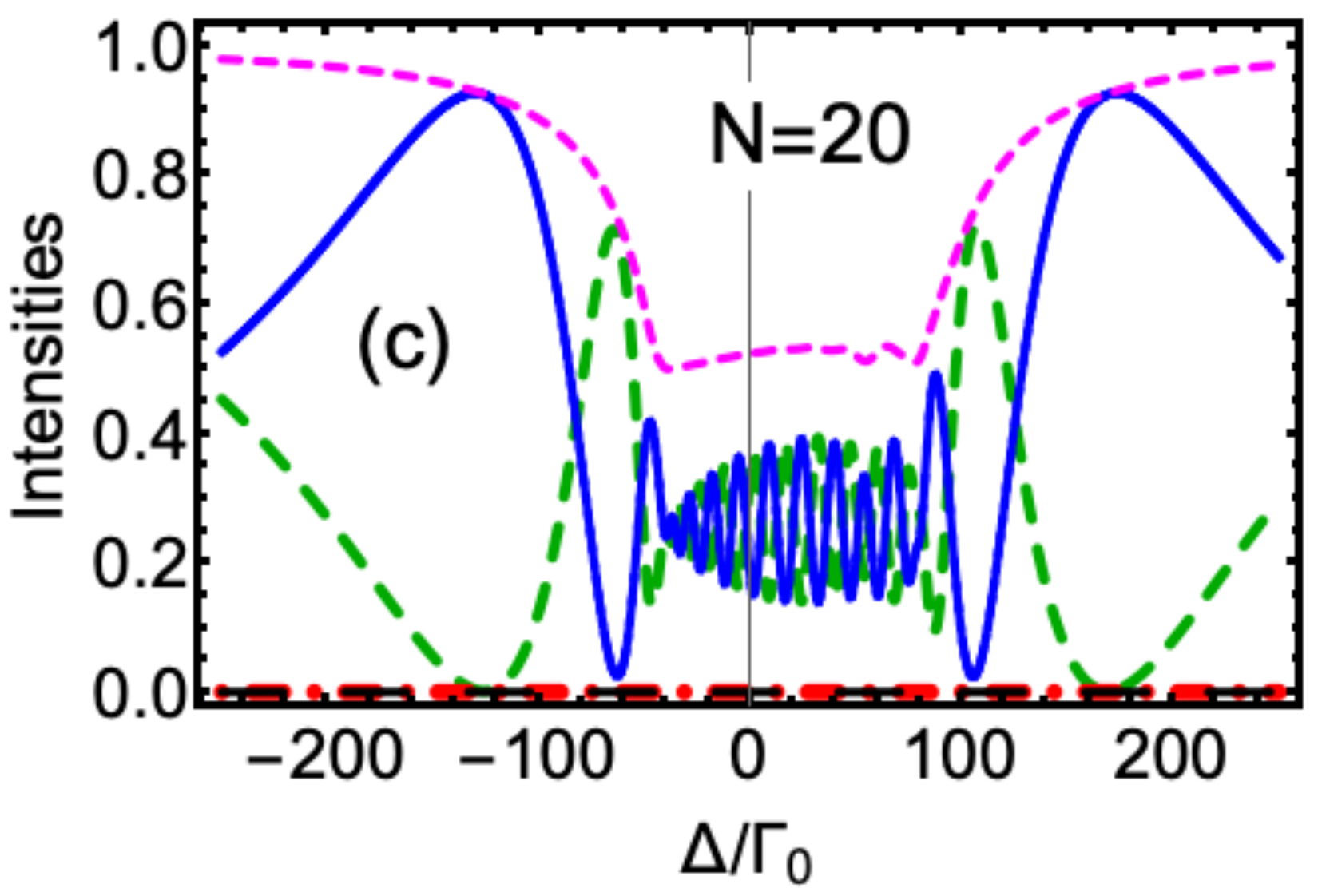} 
    \hspace{12mm}\includegraphics[width=2.5in, height=1.7in]{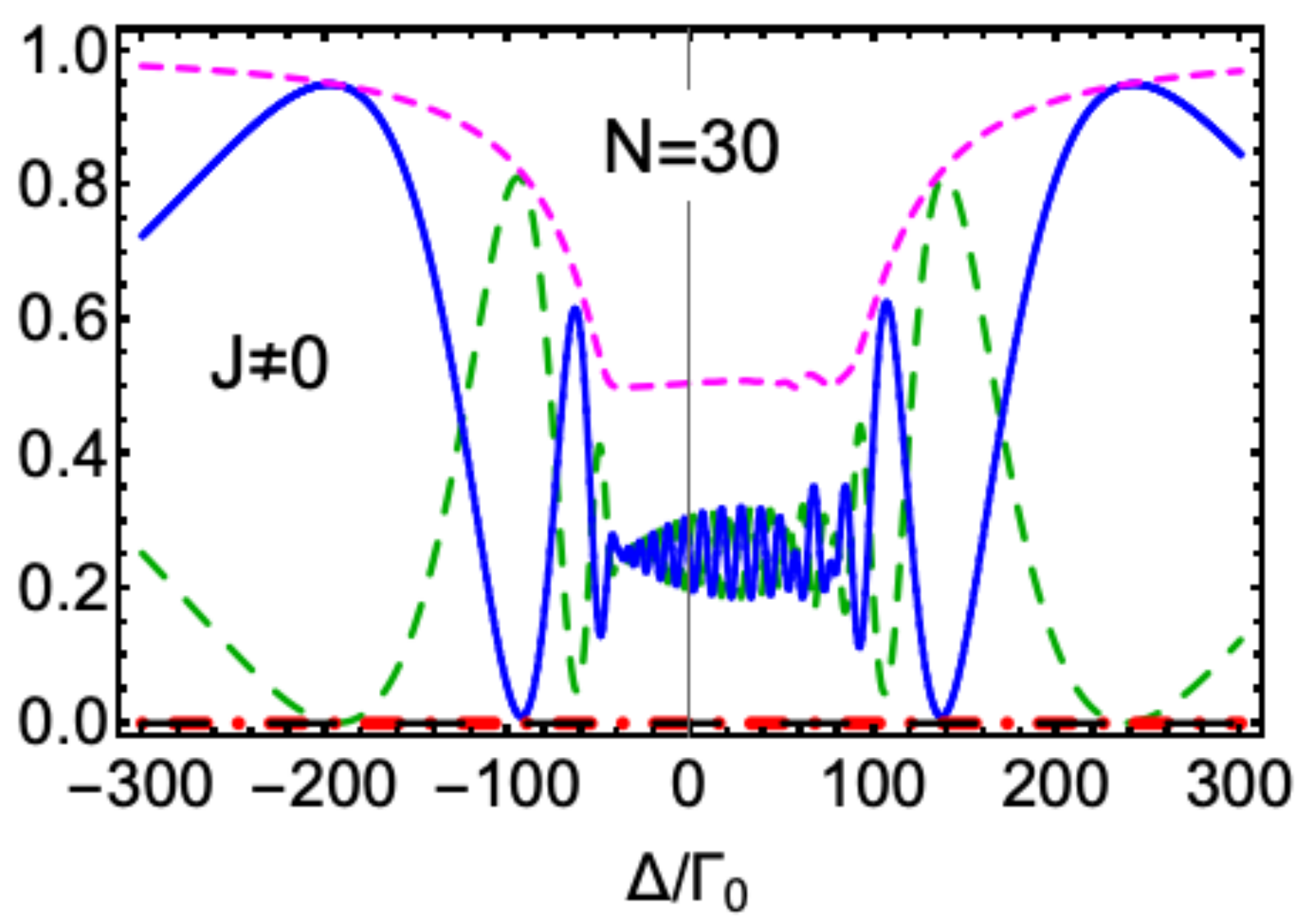} 
  \end{tabular}
\captionsetup{
  format=plain,
  margin=0.1em,
  justification=raggedright,
  singlelinecheck=false
}
 \caption{(Color online) The single-photon transmission \& reflection intensity from the bottom waveguide and leftward \& rightward rectification intensity from the upper waveguide for a chain of (a) five (b) ten (c) twenty and (d) thirty identical dipole-dipole interacting QEs. The separation between any two consecutive emitters in all plots is fixed to $\lambda_{sp}/20$ while $\gamma=6.86\Gamma_{0}$ and $\Gamma=11.03\Gamma_0$. }\label{Fig6}
\end{figure*}

\subsection{Routing improvement due to collective effects of multiple QEs}
Recently it has been reported that the many-emitter waveguide QED setups can outperform single-emitter waveguide QED in performing certain quantum information tasks. For instance, Ryan et al. have shown that collective effects arising from a 10-15 QE chain can result in a near-perfect chiral light-matter interaction \cite{jones2020collectively}. Mahmoodian et al. reported that strongly correlated photon emission can be achieved in optically dense emitter ensembles chirally coupled to waveguides even with weak coupling strengths \cite{mahmoodian2018strongly}. And Mukhopadhyay et al. have discussed the possibility of achieving perfect transparency in a chain of even number of non-identical QEs coupled to a waveguide with separation equal to a half-integral multiple of resonant wavelength \cite{mukhopadhyay2020transparency}. 

Motivated by the aforementioned studies, we now discuss the impact of the collective response of multiple QEs on the photon routing scheme. We find that with more than three QEs analytic expressions become intractable therefore we show numerical results. In Fig.~\ref{Fig6} we present the routing intensities for a moderately sized chain of $30$ QEs. From Fig.~\ref{Fig6}(a) we find that with five QEs both $T=|t_5|^2$ and $\widetilde{T}=|\widetilde{t}_5|^2$ intensities split into four peaks around resonance. Similar to the two QEs cases presented in Fig.~\ref{Fig4}(d), we find that the DDI results in an asymmetrical spectrum. Although with five QEs highest value of $\widetilde{T}$ can reach as high as 72\% at a far detuned value of $\Delta=85.33\Gamma_{0}$. 

\begin{figure}
\centering
   \includegraphics[width=3.25in, height=2.1in]{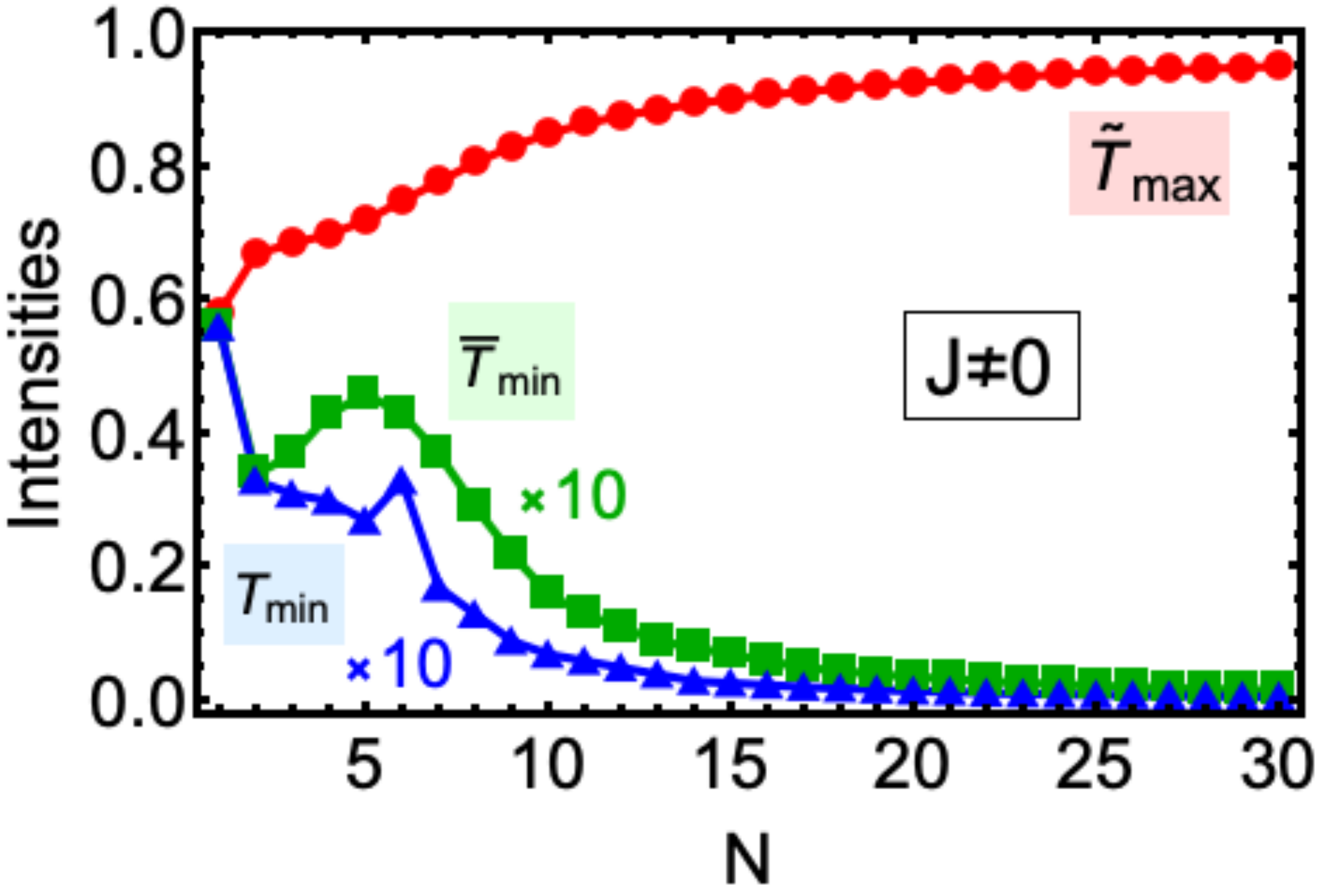} 
\captionsetup{
  format=plain,
  margin=0.1em,
  justification=raggedright,
  singlelinecheck=false
}
 \caption{(Color online) Routing efficiency enhancement as a function of emitter number $N$. $\widetilde{T}_{max}$ represents maximum rightward rectification, $T_{min}$ displays the minimum value of transmission from the lower waveguide, while $\overline{T}_{min}$ is the lower waveguide transmission corresponding to the frequency where maximum $\widetilde{T}_{max}$ has been recorded. Both $T_{min}$ and $\overline{T}_{min}$ have been magnified ten times to fit the scale of the plot. The rest of the parameters are the same as used in Fig.~\ref{Fig6}.}\label{Fig7}
\end{figure}

From Fig.~\ref{Fig6}(b), (c), and (d) we notice that as the number of QEs increases from 5 to 30 the number of near resonance peaks in $\widetilde{T}$ (and $T$) form an asymmetric envelope. Additionally, from $N=10$ onward we observe that the $\widetilde{T}$ spectra exhibit two maximum peaks that appear on both sides of the envelope. For $N=10$, $N=20$, and $N=30$ these peaks appear on the positive detuning axis at $\Delta=108.36\Gamma_0$, $173.33\Gamma_0$, and $241.53\Gamma_0$ with respective $\widetilde{T}$ values of $0.850$, $0.926$, and $0.951$. Interestingly these peaks, which shift to higher $\Delta$ values with longer emitter chains, not only result in a higher rightward rectification but also result in smaller lower waveguide transmission. Notably, such behavior is contrary to a single waveguide QED case where a far-off resonant photon transmits the system without interacting with the emitters \cite{mirza2017chirality}. 

To further emphasize the improvement in routing with longer QE chains, in Fig.~\ref{Fig7} we report emitter number $N$ dependence of maximum rightward rectification $\widetilde{T}_{max}$, minimum lower waveguide transmission $T_{min}$ and lower waveguide transmission $\overline{T}_{min}$ at the same frequency where $\widetilde{T}_{max}$ is achieved. We remark that as $N\longrightarrow 30$, $\widetilde{T}_{max}$ reaches $95\%$ with both $T_{min}$ and $\overline{T}_{min}$ diminishing to almost null values. This behavior demonstrates that the collective effects arising from infinitely long-ranged DDI among $30$ QEs protect the routing scheme from spontaneous emission loss. Here it is worthwhile to point out that the net loss to the environmental degrees of freedom (nonwaveguide modes) can be calculated from the relation $1-T-\widetilde{T}$ for the present chiral setting \cite{asenjo2017exponential}. We indicate that this loss takes almost $5\%$ value for $N=30$ case at the $\Delta$ values where maximum $\widetilde{T}$ is observed (for instance at $\Delta=241.53\Gamma_0$ as shown in Fig.~\ref{Fig6}(d)).


\section{Conclusions}
In encapsulation, we theoretically studied the problem of improving routing efficiency of single photons in a four-port device consisting of a chain of dipole-dipole interacting QEs chirally coupled to a double waveguide ladder in the presence of spontaneous emission. For the case of two QEs, we noticed in the absence of spontaneous emission that the strong waveguide mediated interactions (quantified through the parameter $\Gamma$) leads to splitting of the spectra into two symmetric peaks. The peak separation depends on the $\Gamma$ value. The inclusion of a DDI opens a direct back reflection channel even when both waveguides are chiral. We found that this DDI channel causes an asymmetry in the peak heights and shifts the peak locations. For more realistic scenarios, when spontaneous emission is incorporated, it drastically reduces the peak intensities. However, from the routing point of view, we deduced that two QEs with strong DDI (in QDs coupled to Ag nanowire platforms), can enhance the routing efficiency up to $9\%$ as compared to the single-emitter case when routing was $58\%$ efficient with the same spontaneous emission loss. For the further improvement in the routing scheme, we extended to many-emitter case and numerically found that for a $30$ noisy (i.e. $\gamma\neq 0$) QE chain, the collective effects arising due to long-ranged DDI can redirect the photons from bottom to upper waveguide with $95\%$ efficiency by tuning to far-detuned frequency values. With the current flourishing research activity in light-matter interfaces based on quantum nanophotonics, our results may find applications in designing reliable quantum networks and quantum communication protocols.


\acknowledgments
The authors would like to thank Ana Asenjo-Garcia for useful discussions and Zibo Wang for suggestions on the numerical code. This work is supported by the Miami University College of Arts \& Science and Physics Department start-up funding.


\bibliographystyle{ieeetr}
\bibliography{paper}
\end{document}